\documentclass[fleqn,usenatbib]{mnras}

\usepackage{newtxtext,newtxmath}
\usepackage[T1]{fontenc}

\DeclareRobustCommand{\VAN}[3]{#2}
\let\VANthebibliography\thebibliography
\def\thebibliography{\DeclareRobustCommand{\VAN}[3]{##3}\VANthebibliography}

\usepackage{graphicx}	% Including figure files
\usepackage{amsmath}	% Advanced maths commands
\title[]{Chromosome maps of Globular Clusters from wide-field ground-based photometry}

\author[S.\,Jang et al.]{
S.\,Jang,$^{1,2}$\thanks{E-mail: sohee.jang@unipd.it}
A.\,P.\,Milone,$^{1,3}$
M.\,V.\,Legnardi,$^{1}$
A.\,F.\,Marino,$^{3,4}$
A.\,Mastrobuono-Battisti,$^{5,6}$
E.\,Dondoglio,$^{1}$\newauthor
E.\,P.\,Lagioia,$^{1}$
L.\,Casagrande,$^{7,8}$
M.\,Carlos,$^{1}$
A.\,Mohandasan,$^{1}$
G.\,Cordoni,$^{1}$
E.\,Bortolan,$^{1}$
and Y.-W.\,Lee$^{2}$
\\
$^{1}$Dipartimento di Fisica e Astronomia ``Galileo Galilei'', Universit\`{a} di Padova, Vicolo dell'Osservatorio 3, I-35122, Padua, Italy\\
$^{2}$Center for Galaxy Evolution Research and Department of Astronomy,Yonsei University, Seoul 03722, Korea\\
$^{3}$Istituto Nazionale di Astrofisica - Osservatorio Astronomico di Padova, Vicolo dell'Osservatorio 5, IT-35122, Padua, Italy\\
$^{4}$Istituto Nazionale di Astrofisica - Osservatorio Astrofisico di Arcetri, Largo Enrico Fermi, 5, Firenze, IT-50125\\
$^{5}$GEPI, Observatoire de Paris, PSL Research University, CNRS, Place Jules Janssen, F-92190 Meudon, France\\
$^{6}$Department of Astronomy and Theoretical Physics, Lund Observatory, Box 43, SE-221 00 Lund, Sweden\\
$^{7}$Research School of Astronomy and Astrophysics, The Australian National University, Canberra, ACT 2611, Australia\\
$^{8}$ARC Centre of Excellence for All Sky Astrophysics in 3 Dimensions (ASTRO 3D), Australia
}

\date{Accepted XXX. Received YYY; in original form ZZZ}

\pubyear{2015}

\begin{document}
\label{firstpage}
\pagerange{\pageref{firstpage}--\pageref{lastpage}}
\maketitle

\begin{abstract}
{\it  Hubble Space Telescope} ({\it HST}) photometry is providing an extensive analysis of globular clusters (GCs). In particular, the pseudo two-colour diagram dubbed 'chromosome map (ChM)' allowed to detect and characterize their multiple populations with unprecedented detail. The main limitation of these studies is the small field of view of {\it HST}, which makes it challenging to investigate some important aspects of the multiple populations, such as their spatial distributions and the internal kinematics in the outermost cluster regions.
To overcome this limitation, we analyse state-of-art wide-field photometry of 43 GCs obtained from ground-based facilities. We derived high-resolution reddening maps and corrected the photometry for differential reddening when needed.
We use photometry in the $U, B$, and $I$ bands  to introduce the $\Delta c_{\rm U,B,I}$ vs.\,$\Delta_{\rm B,I}$ ChM of red-giant branch (RGB) and asymptotic-giant branch (AGB) stars. We demonstrate that this ChM, which is built with  wide-band ground-based photometry, is an efficient tool to identify first- and second-generation stars (1G and 2G) over a wide field of view. To illustrate its potential, we derive the radial distribution of multiple populations in NGC\,288 and infer their chemical composition. 
We present the ChMs of RGB stars in 29 GCs and detect a significant degree of variety. The fraction of 1G and 2G stars, the number of subpopulations, and the extension of the ChMs significantly change from one cluster to another. Moreover, the metal-poor and metal-rich stars of Type\,II GCs define distinct sequences in the ChM. We confirm the presence of extended 1G sequences. 

\end{abstract}


\begin{figure*}
\begin{center}
	\includegraphics[height=14.5cm,trim={0cm 0cm 0cm 0cm}]{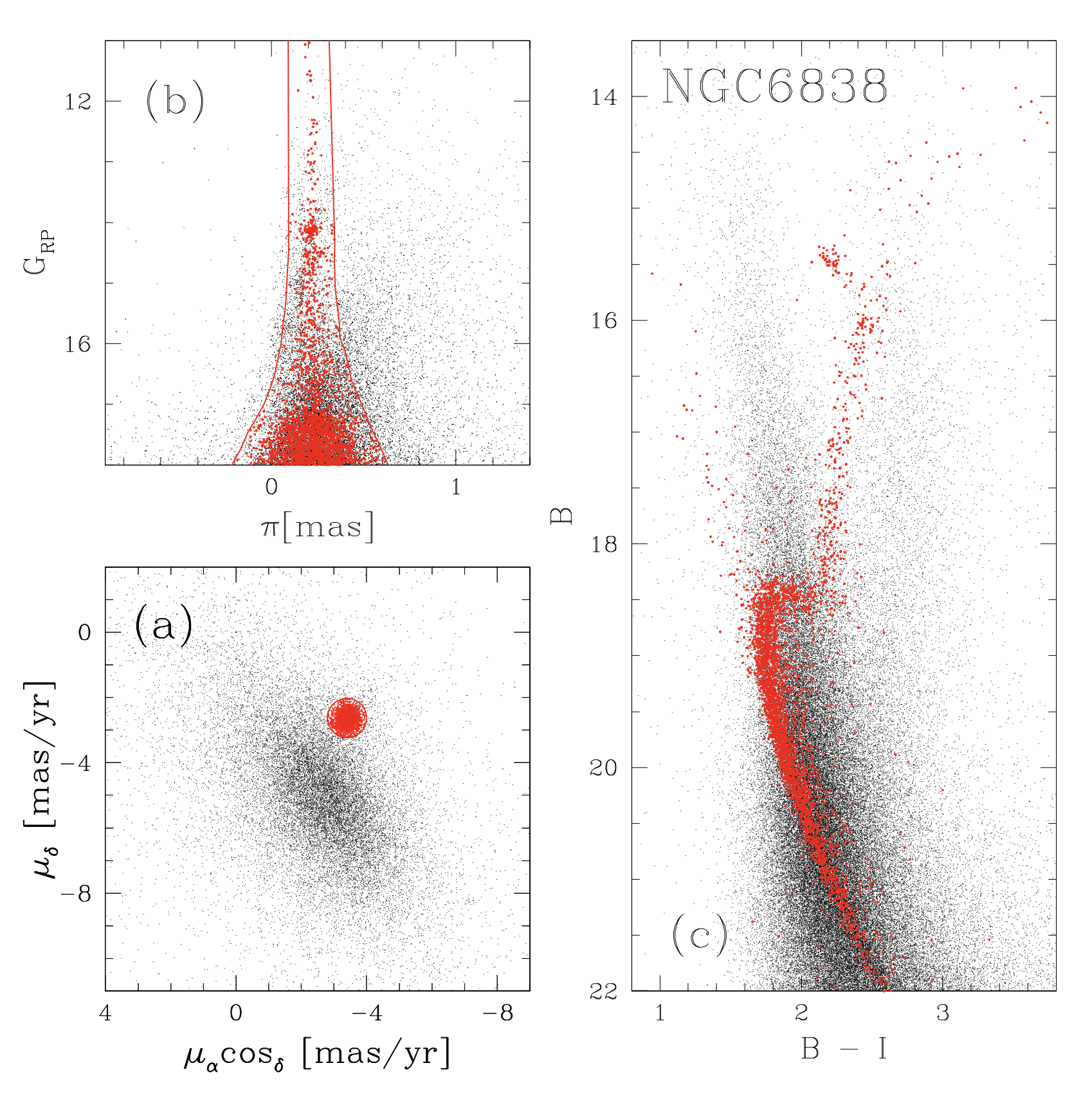}
    \caption{This figure illustrates the procedure to select probable members of NGC 6838. Panel (a) shows the VPD of proper motions for stars in the cluster field. Panel (b) show $G_{\rm RP}$ as a function of the parallax $\pi$ and the two lines drawn by eye that encloses probable cluster members. The B vs. B-I CMD is illustrated in panel (c). Cluster members selected are represented with red symbols. }
    \label{fig:selection}
\end{center}
\end{figure*}
%%%%%%%%%%%%%%%%% BODY OF PAPER %%%%%%%%%%%%%%%%%%

\section{Introduction}
It is now well established that most globular clusters (GCs) host two main groups of stars: a first stellar population (hereafter 1G) with chemical composition that resembles halo field stars, and a second population (2G) of stars enriched in helium, nitrogen and sodium and depleted in carbon and oxygen \citep[see][for recent reviews]{bastian2018a, gratton2019a,milone2022a}. Multiple populations have been widely investigated during the past decades but their origin still represents one of the most intriguing open issues in stellar astrophysics. According to some scenarios, the multiple populations are the signature of multiple star-formation episodes, where 2G stars originate from material polluted by more-massive 1G stars \citep[e.g.][]{dantona2016a, decressin2007a, demink2009a, denissenkov2014a,jang2014a,jang2015,lee1999,lee2016, renzini2022a}. Other formation scenarios suggest that all GC stars are coeval and that accretion of polluted material onto pre-MS stars is responsible for the chemical composition of 2G stars \citep[e.g.][]{bastian2013a, gieles2018a}. As an alternative, stellar mergers are responsible for multiple populations in GCs \citep[][]{wang2020}.

Photometry, together with spectroscopy, is the main technique to identify and characterize multiple populations in GCs, and various photometric diagrams have been exploited in the past decade to disentangle 1G and 2G stars. 
%by using both {\it Hubble space telescope} ({\it HST}) and ground-based facilities. 
 %Appropriate combinations of ultraviolet and optical photometry resulted to be efficient tools to identify the distinct stellar populations in GCs. 
 %In particular, 
 Numerous studies, based on colour-magnitude diagrams (CMDs), two-colour diagrams, and pseudo CMDs that are derived from suitable combinations of 
  stellar magnitudes %in the F225W, F275W, F336W, F343N, F395N, F410M, F438W, and F814W
   filters of {\it Hubble space telescope} ({\it HST}), have provided an homogeneous and extensive analysis of the multiple-population phenomenon in Galactic and extragalactic GCs \citep[e.g.][]{milone2012b, milone2020a, piotto2015a, niederhofer2017a, lagioia2019a, jang2021a}. 
 
 The pseudo two-colour diagram dubbed 'chromosome map' (ChM) resulted to be one of the most efficient diagnostic tools to identify and characterize multiple populations. %in a large sample of about seventy Galactic and extragalactic GCs.
 The traditional ChM, which is built with a suitable combination of stellar magnitudes in the F275W, F336W (or F343N), F438W, and F814W filters, is very sensitive to the chemical composition of GC stars (mostly on the content on helium, nitrogen, carbon, and oxygen).
  It has allowed us to identify and characterize multiple population phenomenon within about sixty Galactic GCs, with unprecedented detail \citep{milone2015a, milone2017a}. 
  
 More recently, various authors introduced additional ChMs based on {\it HST} photometry. Some ChMs are composed of F336W (or F343N), F438W, and F814W magnitudes alone, thus avoiding the most time-demanding F275W filter \citep[][]{zennaro2019a, larsen2019a, milone2020a}. In addition, ChMs build with the F275W, F280N, F343N, and F373N filters allowed us to disentangle stellar populations with different Mg abundances along the RGB \citep[][]{milone2020b}, whereas ChMs composed of appropriate combinations of optical and near-infrared filters (e.g.\,F606W, F814W, F110W, and F160W) are sensitive to multiple populations with different oxygen abundances among very-low mass stars \citep[][]{milone2017b, milone2022a}. 
  
   %In particular, this study has revealed that the fraction of 2G stars of GCs correlates with the cluster mass, showing that incidence and complexity of the multiple population phenomenon increases with cluster mass. In the study, however, the fractions of 1G stars derived from HST photometry are representative of the center of the cluster alone. Although the field of view of most clusters analyzed encloses the half-light radius \citep[see Table 2 in][]{milone2017a}, the entire fraction of 1G stars might be different from that obtained in the central region. Indeed, the 2G stars in massive cluster is more centrally concentrated than the 1G stars.  
  % 
  
   One of the major limitations is the small field of view of {\it HST} cameras, which restricts most of the studies on Galactic GCs to the innermost cluster regions. 
   Clearly, wide-field photometry is mandatory to extend the investigation of multiple populations to the entire cluster.
%Using the diagnostic tool of ChMs of 59 GCs, built with the $m_{\rm F275W}$ - $m_{\rm F814W}$ colour and the $C_{\rm F275W,F336W,F438W}$ = ($m_{\rm F275W}$ - $m_{\rm F336W}$) - ($m_{\rm F336W}$ - $m_{\rm F438W}$) pseudo-colour, has allowed us to quantify and document the multiple population phenomenon among Galactic GCs, showing that incidence and complexity of the multiple population phenomenon increases with cluster mass.  

%has allowed us to document the multiple populations phenomenon in Galactic GCs, showing that the complexity of the phenomenon is intimately associated with the GC mass.

To overcome the main limitation of {\it HST}, ground-based facilities are used to investigate the multiple stellar populations in Galactic GCs by means of photometric diagrams that are sensitive to the chemical composition of the distinct populations.
 The most-used diagrams involve CMDs and pseudo-CMDs built with Str{\"o}emgren photometry \citep[e.g.][]{grundahl1998a, yong2008a}, photometry from appropriate narrow-band filters \citep[][]{lee2017a}, or appropriate combinations of the $U$, $B$, $V$ and $I$ Johnson-Cousin bands \citep[][]{marino2008a, milone2010a, milone2012a, dondoglio2021a}. 

In this context, photometric diagrams made with the $C_{\rm U,B,I} = (U-B) - (B-I)$ pseudo colour, are efficient tools to disentangle stellar populations with different light-element abundances along the red-giant branch (RGB), whereas CMDs made with $B-I$ colour are sensitive to stellar populations with different helium abundances \citep[e.g.][]{monelli2013, cordoni2020a}.
Although these diagrams are widely used to identify multiple populations in GCs, they are less efficient than ChMs to disentangle the distinct stellar populations of several GCs. 
%In addition, a wide field of view of the ground-based UBVI photometry has been taken advantage of to investigate the radial gradients in the distribution of multiple populations, indicating that the 2G stars are more centrally concentrated than the 1G stars in most GCs \citep{sollima2007,bellini2009,milone2012a,dondoglio2022}.

%= ($m_{\rm F275W}$ - $m_{\rm F336W}$) - ($m_{\rm F336W}$ - $m_{\rm F438W}$)

\citet{hartmann2022a} first derived a ChM from ground-based photometry, by using  the $U$ band together with the narrow-band filters centered around 3780\AA\, and 3950\AA.
In this paper, we introduce a new ChM, which is based on photometry in the $U$, $B$, and $I$  bands, and provides improved identification of multiple populations from ground-based Johnson-Cousin photometry. 
In addition to the filter choice, the correction of the effect of differential reddening on stellar magnitudes is a crucial ingredient to disentangle multiple populations in photometric diagrams \citep[][]{milone2012a}. Hence, we derive and publicly release high-resolution reddening maps.

%To do that, we combine photometry from \citet{stetson2019} and proper motions from Gaia early data release 3 \citep[eDR3][]{gaia2021a} to identify the probable cluster members, derive high-resolution differential-reddening maps, and disentangle the distinct stellar populations in the ChMs.
%To investigate the radial distribution and the global fraction of multiple populations, we exploit the ground-based photometry of Galactic GCs and adopted the ChM that maximizes the separation between stellar populations. The photometry studied in the paper has been corrected for differential reddening that are a important step in identifying multiple sequences from photometry. Finally, we have constructed the ChMs of RGB stars of Galactic GCs built with $C_{\rm U,B,I}$ pseudo-colour and B-I colour.

%To extensively investigate the fraction and radial distribution of the distinct group of stars 1G and 2G over a wide field of view in Galactic GCs, it is first required to construct the chromosome maps of them built with UBVI ground-based photometry, in which the two populations of stars can be clearly separated. As a fundamental stage of this study, we have derived the $\Delta c_{\rm U,B,I}$ versus $\Delta_{B,I}$ ChMs of xx GCs, on which the first and second population of stars are well discriminated. The values of differential reddening are also provided for 17 GCs with significant reddening variations.

The paper is organized as follows. Data and data analysis are  described in Section \ref{sec:data}, where we also investigate the differential reddening and illustrate the reddening maps. %of GCs with significant reddening variations and describe how to correct them. 
 The method to derive the ChM is described in Section\,\ref{sec:chms}, where we also provide some applications on how to characterize multiple populations in NGC\,288 by using the ChM.
 The ChMs of RGB stars in 29 Galactic GCs are presented in Section\,\ref{sec:atlas}, while Section\,\ref{sec:AGB} is dedicated to the ChM along the asymptotic-giant branch (AGB). Summary and conclusions follow in Section\,\ref{sec:conclusion}.

\begin{figure}
\begin{center}
	\includegraphics[height=9.0cm,trim={.5cm 0cm 0cm 0cm}]{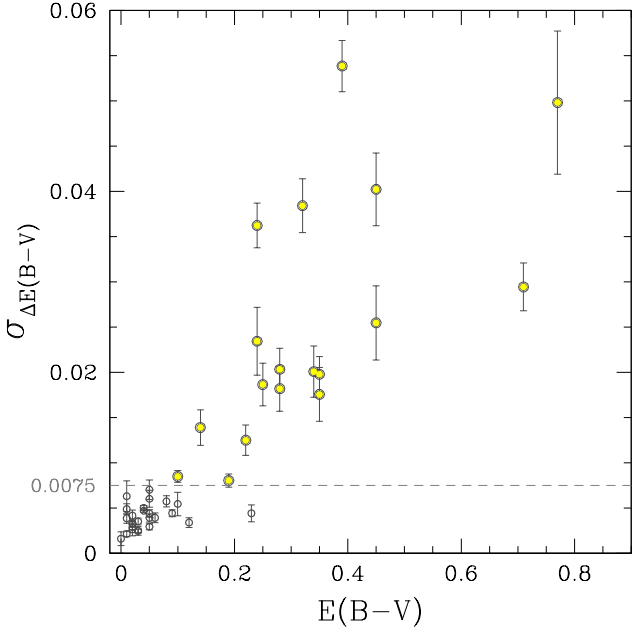}
    \caption{$\sigma_{\Delta E(B-V)}$ against the average reddening from of the host cluster \citep[from the 2010 version of the][catalog]{harris1996a}. 
    $\sigma_{\Delta E(B-V)}$ indicates 1$\sigma$ value of average reddening of each of 100 $\times$ 100 arcsec cells dividing the field of view of a GC. Photometry of 18 GCs with $\sigma_{\Delta E(B-V)}$>0.0075 coloured in yellow have been corrected for the effect of differential reddening in this work.}
    \label{fig:ebv}
\end{center}
\end{figure}

%Although the multiband HST photometry has provided an homogeneous and extensive analysis of the multiple stellar populations in Galactic GCs, its small field of view has prevented us from investigating some of the fundamental aspects of multiple populations, for example the spatial distribution and the entire fraction of the different stellar groups. To overcome this drawback and extend multiple population study over a wide field of view, we have constructed the ChMs of Galactic GCs built based on ground based UBVI photometry.   

%The mF275W - mF814W stellar colours and the CF275W,F336W,F438W pseudo-colours are powerful tools to identify and characterize multiple populations in GCs as they are very sensitive to the stellar abundances of nitrogen and helium, respectively.

%Studies based on the {\it Hubble Space Telescope} (HST) images have demonstrated that appropriate combinations of ultraviolet and optical filter effectively identify multiple stellar populations in the globular clusters. In particular, the pseudo-colour diagram dubbed the chromosome map (ChM) turned to be very effective for identifying distinct sequences  
%%%%%%%%%%%%%%%%%%%%%%%%%%%%%%%%%%%%%%%%%%%%%%%%%%
\begin{figure*}
\begin{center}
	\includegraphics[height=15.5cm,trim={0cm 0cm 0cm 0cm}]{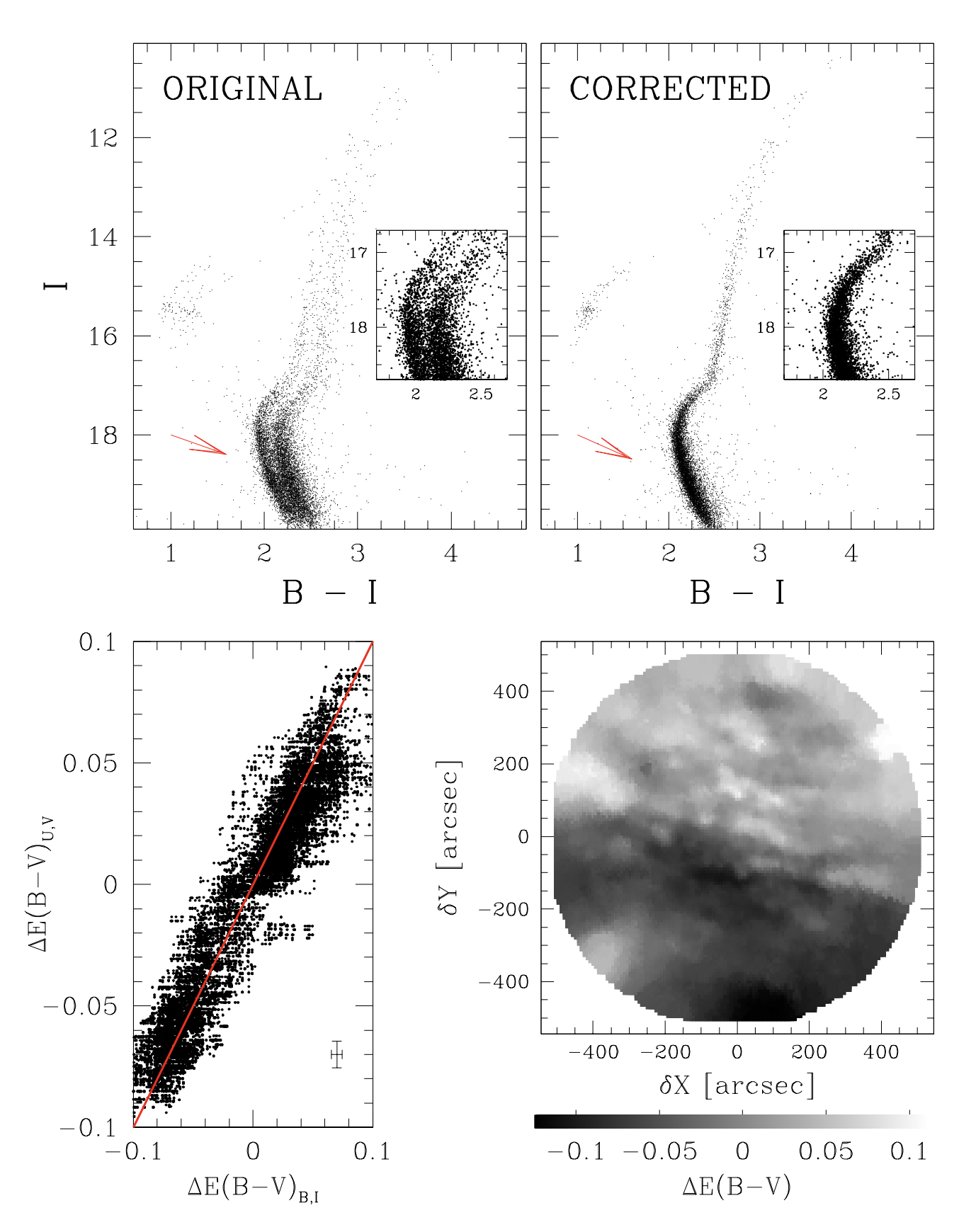}
    \caption{\textit{Upper panels}: comparison of the original $I$ versus $B-I$ CMD of NGC\,4372 (left) and the CMD corrected for differential reddening (right). A zoom of each CMD around the SGB is plotted in the insets. The reddening vectors are indicated with red arrows. \textit{Lower panels}: differential reddening map (right). The levels of grey correspond to different $\Delta$E($B-V$) values as indicated by the scale on the bottom. The left panel compares the differential reddening values inferred from the $I$ versus $B-I$ and the V versus $U-V$ CMD. The red line indicates the perfect agreement.}
    \label{fig:reddening}
\end{center}
\end{figure*}

\section{Data and Data Analysis}\label{sec:data}
To derive differential-reddening maps in the field of view of 43 GCs we use the photometry in the $U$, $B$, $V$, $I$  photometric bands from \citet{stetson2019} and coordinates, proper motions, and parallaxes from the {\it Gaia} early Data Release 3 \citep[][eDR3]{gaia2021a}.
 Photometry has been derived by Peter Stetson from images collected with various ground-based telescopes and by using the methods and the computer programs by \citet{stetson2005} and has been calibrated on the reference system by \citet{landolt1992}. Details on the data and the data analysis are provided by \citet{stetson2019}
 \footnote{ The sample studied by \citet[][]{stetson2019} comprises 48 GCs.
 We excluded from our analysis five clusters, namely $\omega$\,Centauri, because of the extreme complexity of its multiple populations, and E\,3, Palomar\,14, NGC\,5694, and NGC\,5824, due to the small number of cluster members with high-quality photometry and proper motions.}.

%To unambiguously identify multiple sequences from the photometry, it is required to correct the differential reddening and to eliminate field stars from cluster members. To do this, we combined high-precision ground-based photometry of stars with {\it Gaia} Data Release 2 (DR2) proper motion \citep{GaiaDR2}. Photometry in U, B, V, I bands has been derived by Peter Stetson from images collected with various facilities and by using the methods and the computer programs by \citet{stetson2005} and \citet{stetson2019}. Photometry has been calibrated on the reference system by \citep{landolt1992}. Details on the data and the data analysis are provided by \citet{stetson2005}, \citet{monelli2013}, \citet{stetson2019} and references therein. 

\subsection{Selection of cluster members}
To unambiguously identify multiple populations from the photometry, we need to select a sample of  well-measured stars and separate field stars from cluster members. To do this, we combined the ground-based stellar photometry with proper motions and parallaxes from \cite{gaia2021a}.% We downloaded {\it Gaia} DR2 astrometry, photometry, proper motions and parallaxes of stars within the cluster radius, and combined them with the ground-based photometry. 
 We selected the bulk of cluster members by following the procedure by \citet{cordoni2018}, which is illustrated in Figure\,\ref{fig:selection} for NGC\,6838. 
%We selected cluster members for each GC by following the procedure given in \citep{cordoni2018,milone2018a}. 
% The procedure to identify probable cluster members is illustrated in Figure \ref{fig:selection} for the cluster NGC 6838.

 In a nutshell, 
 %first identify a sample of stars with accurate astrometric measurements by using the Renormalized Unit Weight Error (RUWE) parameter  \citep{lindegren2018}. 
 %APM I would add a few words to describe how we selected the well-measured stars from RUWE.
 we first selected stars with cluster-like kinematics, based on their position in the vector-point diagram (VPD) of proper motions. As an example, the probable members of NGC\,6838 are clusters around ($\mu_{\rm \alpha}$cos$\delta$, $\mu_{\rm\delta}$) $\sim$  ($-$3.4,$-$2.6) in the VPD of Figure\,\ref{fig:selection}a. Hence, we draw by hand the red circle to separate bona-fides cluster members from field stars. 
  %We then analyze the vector-point diagram (VPD) of stellar proper motion and, as shown in panel (a), find that the cluster members are clustered around ($\mu_{\rm \alpha}$ cos $\delta$, $\mu_{\rm \delta}$) $\sim$  (-3.4,-2.6). We draw a circle with the coordinate as its center in the VPD that includes the majority of cluster members. Stars put outside of the circle are excluded from the sample of probable cluster members. 
  In addition, we select stars with cluster-like parallaxes, $\pi$, based on their position in the 
   $G_{\rm RP}$ vs.\,$\pi$ plane. As shown in Figure\,\ref{fig:selection}b, the two red lines that we draw by eye enclose the majority of NGC\,6838 stars.
   %APM stars with field-like proper motions should be coloured black in panel b.
   % Panel (b) shows $G_{\rm RP}$ as a function of the parallaxes $\pi$. we draw by eye two lines that encloses probable cluster members. All the stars that lie outside the two lines on the diagram are eliminated from the sample of cluster members.
   The selected well-measured cluster members are represented with red symbols in the $B$ versus $B-I$ CMD of Figure\,\ref{fig:selection}c. 
 %We refer to \citet{cordoni2018} for a detailed description of the procedure. 
 We will use these stars to derive high-resolution reddening maps and to investigate multiple stellar populations.

\subsection{Differential Reddening}\label{sec:dr}
%Foreground dust is not uniformly distributed with position in the cluster, which causes a variation of the reddening, resulting in a non-intrinsic broadening on the CMD. In particular, differential reddening effects is much larger for some cluster, so that an appropriate correction is required for improved analysis of the photometry. Here the U, B, V, I photometry of the selected cluster members has been corrected for differential reddening by following the recipe described in \citet{milone2012a}. To determine the direction of the reddening vector, we used the extinction rate of UBVI bands provided by \citet{schlegel1998}. We refer the reader to the paper by \citet{milone2012a} for further details. 

The distribution of foreground interstellar gas and dust in the directions of the studied GCs is rarely uniform. As a consequence, the amount of reddening suffered by each cluster can vary from one star to another. 
%Clearly, the amount of differential reddening changes from one cluster to another.

To quantify the amount of reddening variation across the field of view of each cluster and to correct the photometry for the effect of differential reddening, we adapted the method by \citet{milone2012a} to the photometric catalogs by \citet{stetson2019}. 
We used the $I$ vs.\,$B-I$ and $V$ vs.\,$U-V$ CMDs to derive two distinct determinations of the amount of differential reddening suffered by each star in each cluster ($\Delta$E($B-V$)$_{B,I}$, and $\Delta$E($B-V$)$_{U,V}$, respectively). 
 The values of $\Delta$E($B-V$)$_{B,I}$ and $\Delta$E($B-V$)$_{U,V}$ are then averaged together to derive improved values of differential-reddening.
%%%%%%%%%%%%%%%%%%%%%%%%%%%
% We provide in Table \ref{tab:DR} the value of 68.27$^{\rm th}$ percentile of $\Delta E(B-V)_{\rm B-I}$ - $\Delta E(B-V)_{\rm U-V}$ to quantify the effectiveness of the differential-reddening correction we have applied \footnote{We excluded from our analysis five clusters studied by \citet[][]{stetson2019}, namely $\omega$\,Centauri, because of the extreme complexity of its multiple populations, and E\,3, Pal\,14, NGC\,5694, and NGC\,5824, due to the small number of cluster members with high-quality photometry and proper motions.}.
 %APM one cluster is missing ?
 % Table 2 includes 43 GCs, we excluded 4 ---> 47 GCs NOT 48 ?
 % We have excluded from our analysis NGC\,5139 for which a poor correlation between the two $\Delta E($B-V$)$ values exhibited, $\sigma_{\rm (\Delta E(B-V)_{\rm B-I} - \Delta E(B-V)_{\rm U-V})}$ = 0.025 mag, together with NGC\,5824, E\,3, and Pal\,14 which have a too-small number of selected stars to analyze. 
 % E3 ?, NGC 5139, NGC 5824, E 3, and Pal 14,

We first selected a sample of reference stars, including bright-MS, sub-giant branch (SGB), and RGB cluster members, and calculated the corresponding fiducial line in the CMD. 
 Then, we used the absorption coefficients by \citet{schlegel1998} to determine the direction of the reddening and  measure the colour and magnitude distance of each reference star from the fiducial along the reddening direction, $d$.
 The values of $d$ have been used for two main  purposes: i) deriving low-resolution reddening maps, which allow to homogeneously compare the variation of reddening in the direction of each cluster, and ii) correct the photometry for the effect of differential reddening. 
 
To quantify the amount of differential reddening in the direction of each cluster, we divided the field of view into 100$\times$100 arcsec cells. 
We estimated the amount of differential reddening of each cell by calculating the median value of $d$ for the reference stars in the cell. This quantity is converted into a differential reddening value by using the coefficients by \citet{schlegel1998}. 
 We provide in Table\,\ref{tab:DR} the 68.27$^{\rm th}$ percentile of the differential-reddening values, $\sigma_{\Delta E(B-V)}$, the difference between the 98$^{\rm th}$ and the second percentile of the differential-reddening distributions $\Delta$E(B$-$V)$_{98\%-2\%}$, and the maximum radial distance used to derive differential reddening, R$_{\rm max}$.  
 The uncertainties associated with these values are calculated with a bootstrap analysis based on 1,000 samples created by a random sampling with replacement of the values of differential reddening. For each extraction we derived the values of $\sigma_{\Delta E(B-V)}$ and $\Delta$E(B$-$V)$_{98\%-2\%}$ by using the procedure described above.
  We considered as our best uncertainties estimates, the value of the random mean scatter of the 1,000 determinations of $\sigma_{\Delta E(B-V)}$ and $\Delta$E(B$-$V)$_{98\%-2\%}$.

 %and .
 
 The values of $\sigma_{\Delta E(B-V)}$ range from less than 0.005 mag to more than 0.050 mag, whereas $\Delta$E(B$-$V)$_{98\%-2\%}$ varies from less than 0.01 mag to $\sim$0.19 mag. As shown in Figure\,\ref{fig:ebv}, $\sigma_{\Delta E(B-V)}$ correlates with the average reddening of the host cluster \citep[from the 2010 version of the][catalog]{harris1996a}.

 % descrizione risultati tabella
 % descrizione figura 4
 The main effect of differential reddening on the photometry is a colour and magnitude broadening of the various sequences of the CMD. 
 To correct photometry for this effect, in 18 GCs with $\sigma_{\Delta E(B-V)}$>0.0075 mag (yellow dots in Figure\,\ref{fig:ebv}), we derived the amount of differential reddening associated with all stars in the catalog.
 For each star, we selected the sample of $35 - 75$ neighboring reference stars and calculated the median distance along the reddening line. We imposed that the maximum radius of the region that includes the reference stars is smaller than 2 arcmin. Hence, the value of  R$_{\rm max}$, is fixed on the basis of this criterion. The only exception is Palomar\,11 where, due to the small number of stars, we used a threshold of 2.5 arcmin.
  We excluded each reference star from the determination of its own differential reddening. 
 The uncertainty associated with differential-reddening determination is estimated as the random mean scatter of the distance values divided by the square root of $N-1$. See \citet[][]{milone2012a} for details. 
  %APM. We should indicate the typical values of N.
 
Some results from this procedure are illustrated in Figure\,\ref{fig:reddening} for NGC\,4372.
Upper panels of Figure\,\ref{fig:reddening} compare the original $I$ versus $B -I$ CMD of NGC\,4372 (left) with the CMD corrected for differential reddening (right). The inset of each panel shows a zoom in the SGB region, where the effect of the correction is more evident. 

Lower-right panel of Figure \ref{fig:reddening} shows the resulting differential reddening map, which reveals spatial variations of $\sim 0.19$ mag in $\Delta E(B$-$V)$, over the field of view. %The error on differential-reddening determination, is $\sigma E(B$-$V)_{\rm B,I} \sim$ XXX mag. To investigate whether the adopted colour and magnitude affect the obtained differential reddening, we compare results from different CMDs. We generate two independent CMDs with photometry in four bands, I versus B - I and V versus B - V. 
The lower-left panel of Figure \ref{fig:reddening} compares $\Delta E(B - V)_{\rm B,I}$ and $\Delta E(B - V)_{\rm U,V}$. The average difference between the two differential-reddening determinations is nearly zero (0.0009 mag), but the random median scatter of 0.016 mag, is larger than the typical uncertainty associated with our reddening determinations.
% SH : random median scatter means sigma ?
Indeed, in addition to differential reddening, small inaccuracies of the PSF model and of the sky determination can contribute to the broadening of the sequences in the CMD. Their presence is indicated by the fact that differences between $\Delta E(B - V)_{\rm B,I}$ and $\Delta E(B - V)_{\rm U,V}$ are typically larger than the observational errors \citep[see][for discussion]{anderson2008a, milone2012a}.
The values of the 68.27$^{\rm th}$ percentile of differential-reddening difference $\Delta E(B-V)_{\rm B-I}$ - $\Delta E(B-V)_{\rm U-V}$ are provided in Table\,\ref{tab:DR}. 
%

% Effetto ZP errors. 
 
%Left two panels of Figure\,\ref{fig:cmddr} compare the original $I$ versus $B -I$ CMD (left) with the CMD corrected for differential reddening (right) for NGC\,3201, NGC\,4833, NGC\,5927, and NGC\,5927.  
 
\begin{figure*}
\begin{center}
	\includegraphics[height=20.5cm,trim={0cm -1cm 0cm 0cm}]{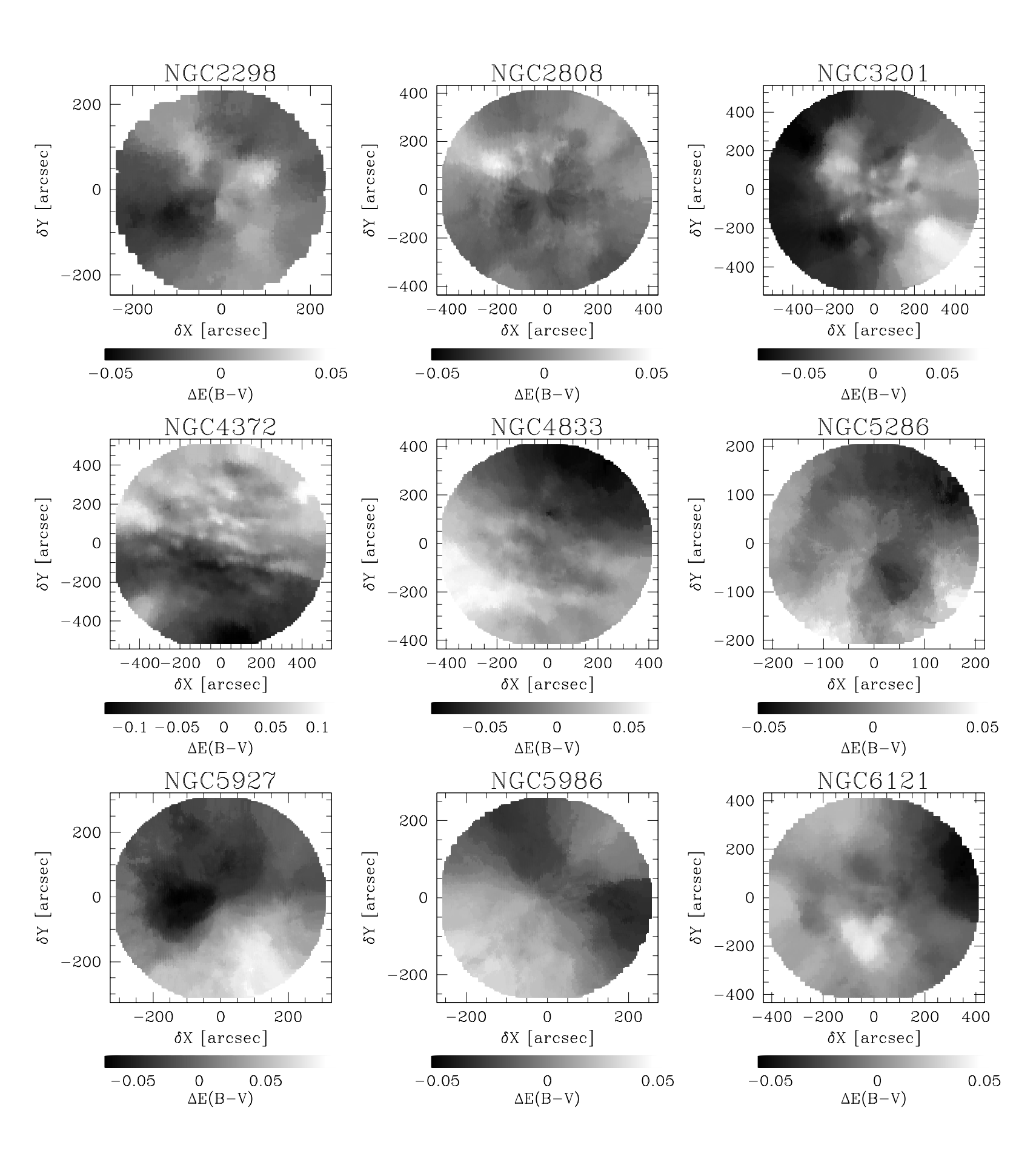}
    \caption{Differential-reddening maps of the regions in front of NGC\,2298, NGC\,2808, NGC\,3201, NGC\,4372, NGC\,4833, NGC\,5286, NGC\,5927, NGC\,5986, and NGC\,6121.}
    \label{fig:drms1}
\end{center}
\end{figure*}

\begin{figure*}
\begin{center}
	\includegraphics[height=20.5cm,trim={0cm -1cm 0cm 0cm}]{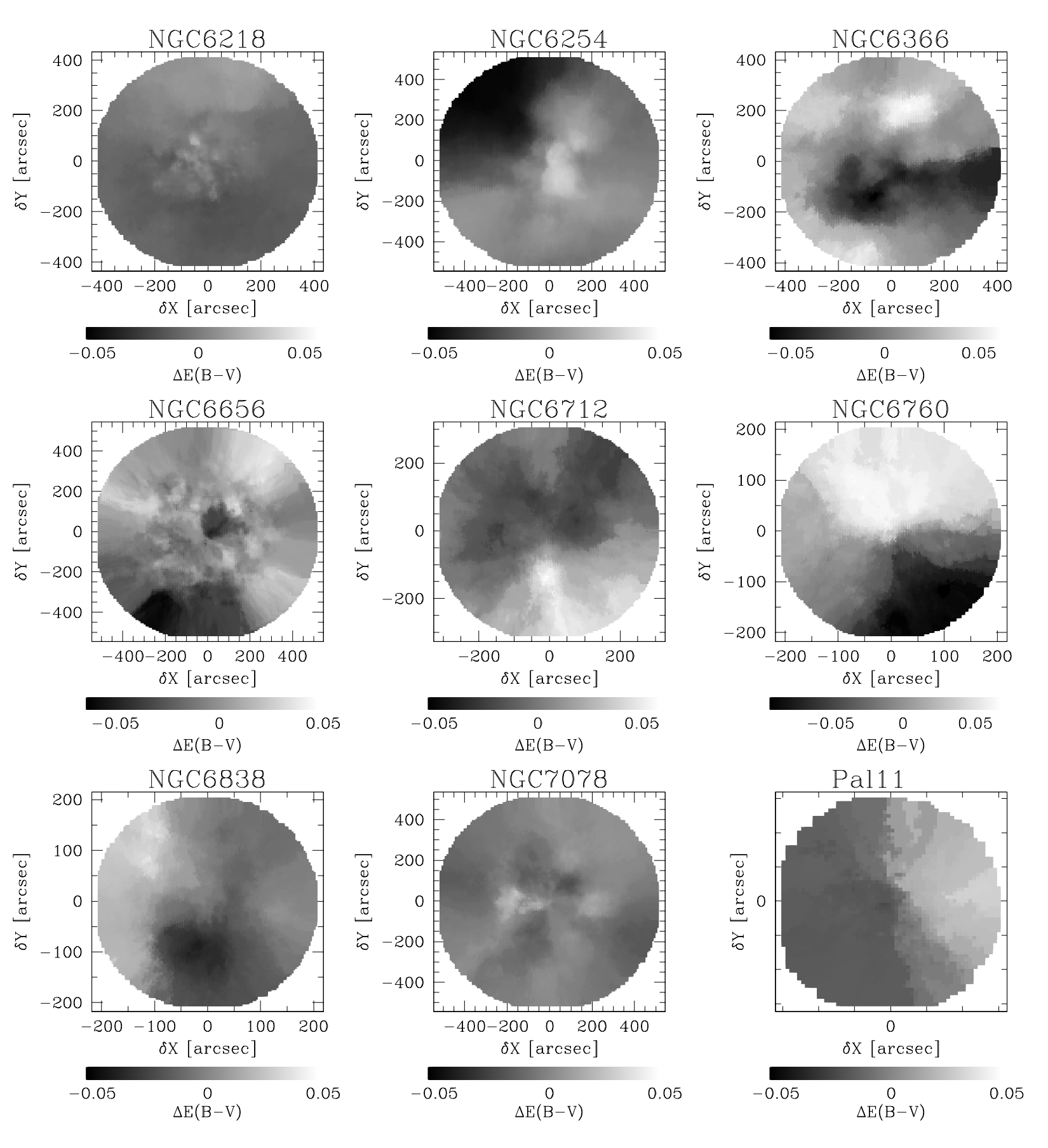}
    \caption{Same as Figure\,\ref{fig:drms1} but for NGC\,6218, NGC\,6254, NGC\,6366, NGC\,6656, NGC\,6712, NGC\,6760, NGC\,6838, NGC\,7078, and Palomar\,11.}
    \label{fig:drms2}
\end{center}
\end{figure*}

The differential-reddening maps for the eighteen GCs with $\sigma_{\Delta E(B-V)}$>0.0075 mag are illustrated in Figure\,\ref{fig:drms1} and Figure\,\ref{fig:drms2} \footnote{The reddening maps are publicly available at this website: http://progetti.dfa.unipd.it/GALFOR/}.

\begin{figure*}
\begin{center}
	\includegraphics[height=17.5cm,trim={0cm 0cm 0cm 0cm}]{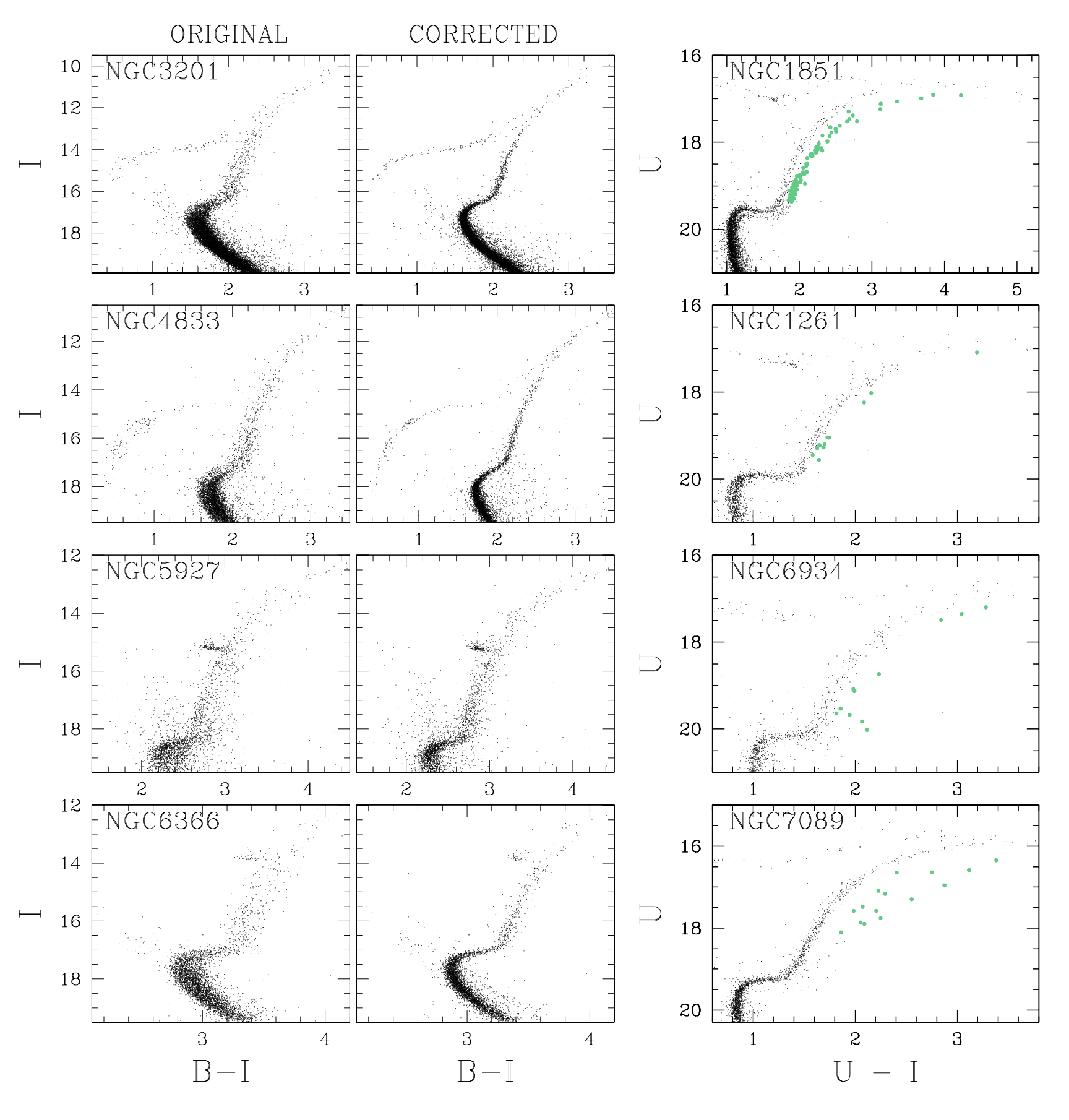}
    \caption{Comparison of the original $I$ versus $B-I$ CMDs of NGC\,3201, NGC\,4833, NGC\,5927, and NGC\,6366 (left panels) with the CMDs corrected
for differential reddening (midle panels). The right panels show the $U$ vs. $U-I$ CMD of the Type\,II GCs NGC\,1851, NGC\,1261, NGC\,6934 and NGC\,7089.
Red-RGB stars are coloured aqua.}
    \label{fig:cmddr}
\end{center}
\end{figure*}

As an example, Figure\,\ref{fig:cmddr} compares the original CMDs of selected cluster members of NGC\,3201, NGC\,4833, NGC\,5927, and NGC\,6366  with the corresponding CMDs corrected for differential reddening. 
 We also show the $U$ vs.\,$U-I$ CMDs of selected cluster members in the Type\,II GCs NGC\,1851, NGC\,1261, NGC\,6934 and NGC\,7089. The blue and red-RGBs are clearly visible and we used aqua colours to highlight red-RGB  stars.

%This procedure has been applied to 48 GCs in the photometric catalogs and among them we provide the values of differential reddening for 17 GCs with significant reddening variation, the 68.27th percentile of the differential reddening distribution $>$ 0.01, NGC 2298, NGC2808, NGC3201, NGC4372, NGC4833, NGC5286, NGC5927, NGC 5986, NGC 6121, NGC 6218, NGC6254, NGC 6366, NGC 6656, NGC 6712, NGC 6760, NGC 6838, and Pal 11, on the GALFOR website\footnote{http://progetti.dfa.unipd.it/GALFOR/}. Figure \ref{fig:drms} shows the resulting differential reddening maps of 9 GCs with the most significant differential reddening variations up to 0.17 mag for NGC 6760.
%This procedure has been applied to the each CMD analysed in this paper, separately. 

\section{Chromosome maps of RGB stars}\label{sec:chms} %of Globular Clusters from wide-field ground-based photometry}

%the colour B - I and the pseudo colour $C_{\rm U,B,I}$ to construct the $\Delta c_{\rm U,B,I}$ versus $\Delta_{\rm B, I}$ pseudo-two-colour diagrams, by following the recipe by \citet{milone2017a} to identify the multiple stellar populations over a wide field of view.
%We have constructed the chromosome map of Galactic clusters built using the colour B - I and the pseudo colour $C_{\rm U,B,I}$ by following the recipe by \citet{milone2017a} to identify the multiple stellar populations over a wide field of view. 
% We refer to the paper by Milone and collaborators for details. In the following subsections, we briefly describe how we measured the intrinsic B-I and $C_{\rm U,B,I}$ RGB width, and how we exploit these two quantities to construct the ChMs of Galactic GCs from ground-based photometry. 

\begin{figure*}
\begin{center}
	\includegraphics[height=14.5cm,trim={0cm 0cm 0cm 2cm}]{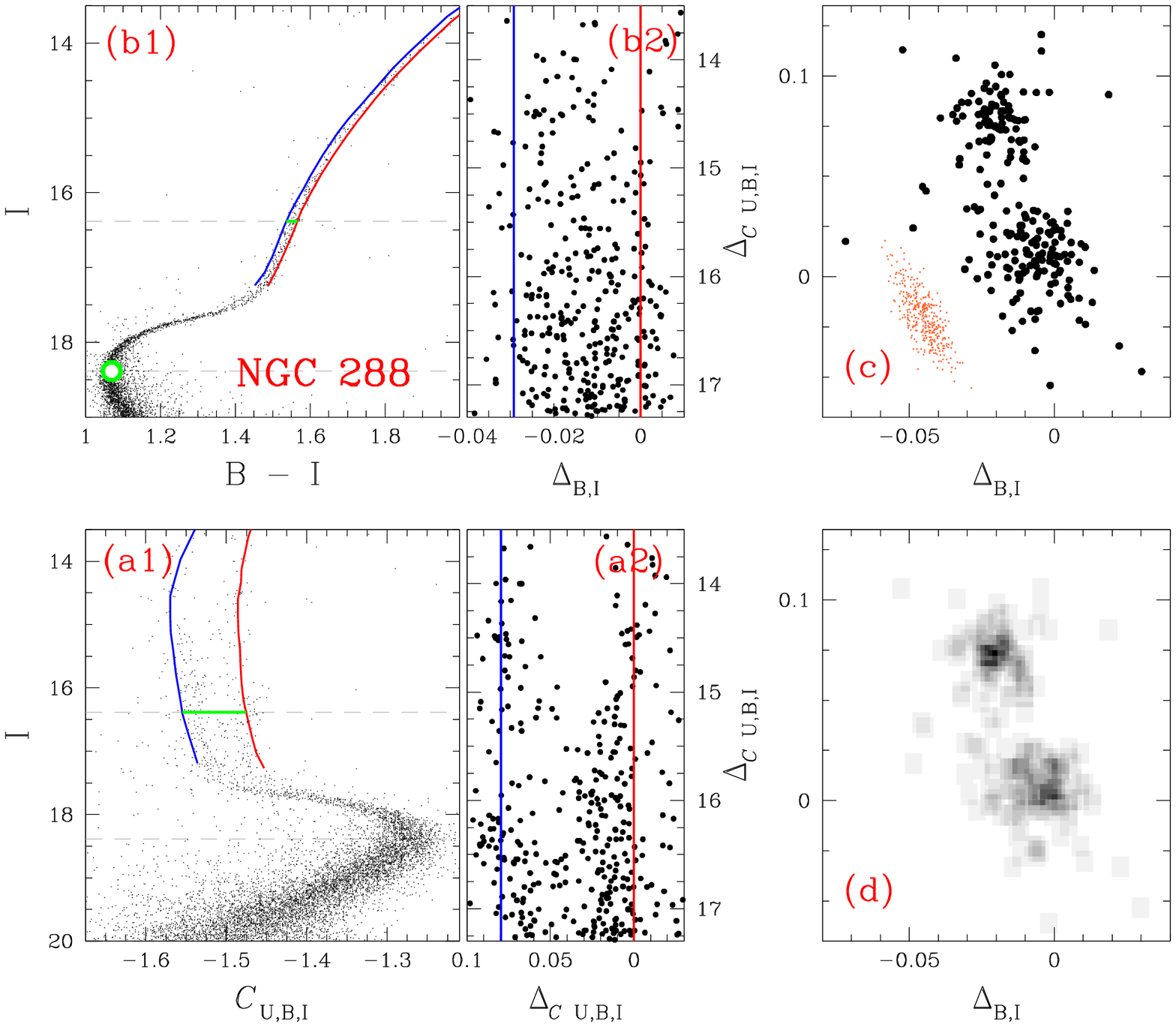}
    \caption{This figure illustrates the procedure to derive the $\Delta c_{\rm U,B,I}$ versus $\Delta_{\rm B,I}$ Chromosome map for the prototypical cluster NGC 288. Panels (a1) and (b1) show the $I$ versus $C_{\rm U,B,I}$ pseudo-CMD and $I$ versus $B-I$ CMD of NGC 288. The green circle in panel (b1) marks the MS turnoff, whereas the two horizontal dotted lines in panels (a1) and (b1) are placed at the magnitude level of the MS turnoff and 2.0 $I$ mag above it. The blue and red lines mark the boundaries of the RGB, while the green segments in the panels (a1) and (b1) indicate the $B-I$ colour and the $C_{\rm U,B,I}$ pseudo-colour separation between the two lines at 2.0 $I$ mag above the MS turnoff. The ‘verticalized’ $I$ versus $\Delta c_{\rm U,B,I}$ and $I$ versus $\Delta_{\rm B,I}$ diagrams for RGB stars are plotted in panels (a2) and (b2), respectively, where the red and blue vertical lines correspond to the RGB boundaries in panels (a1) and (b1) that translate into vertical lines in panel (a2) and (b2). The RGB stars used to construct the chromosome map in panel (c) are represented in panels (a2) and (b2), where $\Delta c_{\rm U,B,I}$ and $\Delta_{\rm B,I}$ are plotted against $I$. Panel (d) shows the Hess diagram for stars in panel (c).}%defined in equations (1) and (2) as explained in the text. }
    \label{fig:procedureChM}
\end{center}
\end{figure*}

To derive the ChM of RGB stars from wide-field ground-based photometry, 
 we combined the $I$ vs.\,$B-I$  CMD and the $I$ vs.\,$C_{\rm U,B,I}$ pseudo CMD of each GC.
 The procedure is illustrated in Figure\,\ref{fig:procedureChM} for NGC\,288, which is used as a test case and is based on the method by \citet[][]{milone2015a, milone2017a}.
 
 The ChM is derived from the $I$ vs.\,$C_{\rm U,B,I}$ pseudo CMD and the $I$ vs.\,$B-I$ CMD plotted in panels (a1) and (b1), respectively. The red and blue fiducial lines superimposed on each diagram are derived by hand and mark the red and blue boundary of the RGB, respectively. 
%We then chemically characterize the multiple populations identified on the chromosome map to justify our identification of 1G and 2G stars as indeed belonging to the first and the second generation. We also confirm whether the identification of 1G and 2G from the $\Delta c_{\rm U,B,I}$ versus $\Delta_{\rm B, I}$ ChM is identical to that shown in the $\Delta_{\rm C_{\rm F275W,F336W,F438W}}$ versus $\Delta_{\rm F275W,F814W}$ ChM.

%The chromosome maps' of the multiple stellar populations based on ground-based photometry over wide fied of view

%The photometric catalogs by Stetson and collaborators have been widely used to investigate multiple populations in GCs \citep[e.g.,][]{milone2012b,milone2018a,monelli2013,marino2016,marino2017a,stetson2019,cordoni2020a,cordoni2020b,dondoglio2020a}. Most of these studies are based on the pseudo colour $C_{\rm U,B,I} = (U-B) - (B-I)$, which is an excellent tool to disentangle stellar populations with different light-element abundance along the RGB. 

%\subsection{Building the chromosome map}
%\label{sub:ChM}

%The colour broadening of the RGB stars wider than expected from the photometric errors provides evidence for the presence of multiple populations in GCs. In this paper, the intrinsic B - I and $C_{\rm U,B,I}$ width of RGB stars are used to construct the chromosome map of each cluster. The procedure to estimate the RGB width in the I versus B - I and $C_{\rm U,B,I}$ plots is illustrated in Figure \ref{fig:procedureChM} for NGC 288 as an example cluster. Due to the small number of the RGB stars, the red and blue boundaries of the RGB on the CMD has been drawn by eye as illustrated in panels (a1) and (b1), in which blue and red lines mark the envelopes of the RGB. The observed RGB width, $W_{C \rm U,B,I}$, has been derived as the difference between the $C_{\rm U,B,I}$ index of the red and blue fiducial lines, calculated 2.0 $I$ mag above the MS turnoff, as represented by the green line in panel (a1) of Figure \ref{fig:procedureChM}. We then have 
The RGB boundaries are used to ‘verticalize’ the two diagrams in a way that each of them translates into a vertical line. Specifically, we defined 
\begin{equation}
    \Delta_{\rm B,I} = W_{\rm B,I}  \frac{X - X_{\rm fiducial R}}{X_{\rm fiducial R} - X_{\rm fiducial B}}
	\label{eq:quadratic}
\end{equation}

\begin{equation}
    \Delta c_{\rm U,B,I} = W_{C \rm U,B,I}  \frac{Y - Y_{\rm fiducial B}}{Y_{\rm fiducial R} - Y_{\rm fiducial B}}
	\label{eq:quadratic2}
\end{equation} 

where $X = B - I$ and $Y = C_{\rm U,B,I}$ and ‘fiducial R’ and ‘fiducial B’ correspond to the red and the blue fiducial lines, respectively. The quantities $W_{\rm B,I}$ and $W_{C \rm U,B,I}$ are indicative of the RGB widths. Their values correspond the colour separation between the RGB boundaries 2.0 $I$ mag above the main-sequence (MS) turnoff.
%as shown in panels (a2) and (b2) of Figure \ref{fig:procedureChM}.

The ‘verticalized’ $I$ versus $\Delta c_{\rm U,B,I}$ and $I$ versus $\Delta_{\rm B,I}$ diagrams for RGB stars are plotted in panels (a2) and (b2) of Figure \ref{fig:procedureChM}, respectively, where the red and blue vertical lines correspond to the RGB boundaries in panels (a1) and (b1). %that translate into vertical lines in panel (a2) and (b2). This is obtained by defining for each star:
 %Thus, $\Delta_{B,I}$ = 0 and $\Delta c_{\rm U,B,I}$ = 0 correspond to stars lying on the corresponding red fiducial line for B - I and blue fiducial for $C_{\rm U,B,I}$ and $\Delta$ quantities represent the colour and pseudo-colour distance from such lines. Note that, as illustrated in Panel (a1), stars likely belonging to the first population are placed along the blue fiducial line on the the pseudo-CMD, contrary to what is shown on I versus B - I CMD.
The $\Delta_{\rm B-I}$  and $\Delta c_{\rm U,B,I}$ pseudo colours are used to derive the ChM plotted in panel (c), where we only included RGB stars with $13.0 < I < 17.5$ mag. As highlighted by the Hess diagram shown in panel (d), the ChM of NGC\,288 reveals two distinct stellar populations clustered around  ($\Delta_{\rm B,I}$,$\Delta c_{\rm U,B,I}$),=(0.0,0.0) mag and ($-$0.02,0.07) mag.

\begin{figure}
\begin{center}
	\includegraphics[height=9.0cm,trim={1cm 0cm 0cm 0cm}]{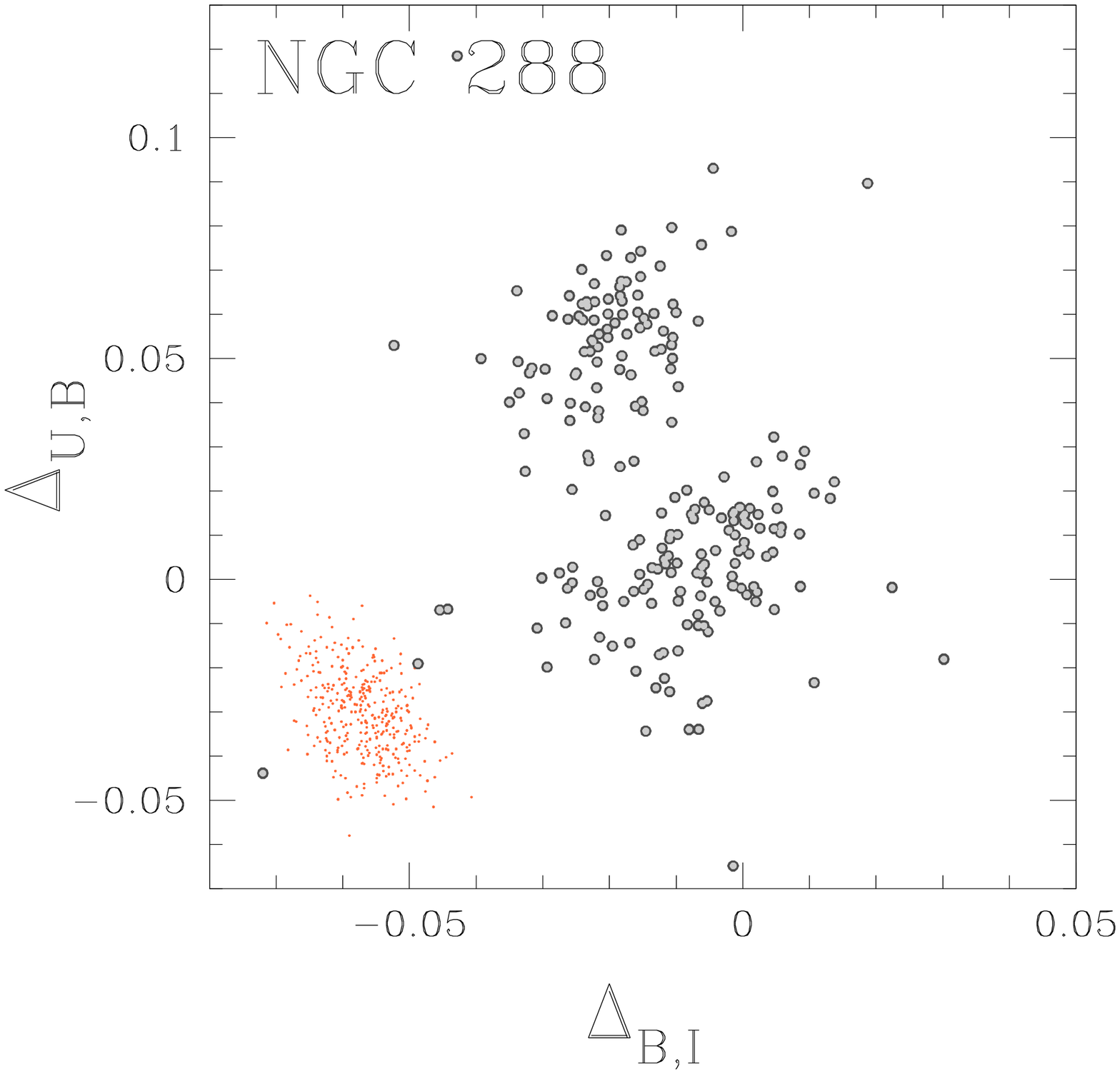}
    \caption{$\Delta_{\rm U,B}$ versus $\Delta_{\rm B,I}$ ChM for RGB stars of NGC\,288.}
    \label{fig:chm_UB}
\end{center}
\end{figure}

For completeness, we show in Figure\,\ref{fig:chm_UB} the $\Delta_{\rm U,B}$ vs.\,$\Delta_{\rm B,I}$ ChM of RGB stars in NGC\,288. This ChM is similar to the ChM introduced in Figure\,\ref{fig:procedureChM}, but is derived from the $I$ vs.\,$U-B$ and $I$ vs.\,$B-I$ CMDs. Here, one group of stars is clustered around the origin of the reference frame, whereas the remaining stars are distributed around ($\Delta_{\rm B,I}$,$\Delta_{\rm U,B}$)$\sim$($-$0.2,0.06) mag. 
 Although for simplicity, this paper is focused on the $\Delta_{\rm B-I}$  vs.\,$\Delta c_{\rm U,B,I}$ diagrams alone, the ChM introduced in  Figure\,\ref{fig:chm_UB}  provides an additional tool to investigate multiple populations in GCs.

Next Section\,\ref{sub:teo} takes advantage of isochrones and synthetic spectra for better understanding the position of stars in the ground-based ChM.
In Section\,\ref{sub:HST} we compare the ChMs of NGC\,288 from {\it HST} and ground-based photometry, while in Sections\,\ref{sub:spectroscopy} and \ref{sub:RD} we illustrate some applications of the ChM of NGC\,288 to constrain the chemical composition of 1G and 2G stars, and investigate their radial distributions.

\subsection{Theoretical interpretation}\label{sub:teo}
%%%%%%%%%%%%%%%%%%%%%%%%%%%%%%%%%%%%%%%%%%%%%%%%%%%%%
To investigate the position of stars with different contents of He, C, N, O, and Fe in the ChM of Type\,I GCs, we started deriving the colours and magnitudes of the Dartmouth  isochrones \citep[][]{dotter2008a} with the same age of 13 Gyr, metallicity, [Fe/H]=$-$1.5, and [$\alpha$/Fe]=0.4, but different chemical compositions.  To do this, we followed the procedure by \citet[][]{milone2018a}, which uses model atmospheres and synthetic spectra. 
 We identified fifteen points along the isochrone, I1, that is representative of 1G stars and extracted their stellar parameters. We calculated two spectra for each combination of effective temperature and gravity, including a reference spectrum that resembles 1G stars and several comparison spectra that share the same chemical composition as 2G stars. We assumed for 1G stars Y=0.246, solar-scaled abundances of carbon and nitrogen, and [O/Fe]=0.4, while the chemical compositions used for 2G stars are listed in Table\,\ref{tab:TEO}. 
 
We used the ATLAS12 and SYNTHE programs \citep[][]{kurucz1970a, kurucz1981a, kurucz2005a, sbordone2004a, sbordone2007a} to derive model atmosphere structures and compute synthetic spectra. When we modified the helium content we accounted for the variation
in effective temperature and gravity predicted by the isochrones by \citet[][]{dotter2008a}. We integrated each spectrum over the bandpasses of the $U, B, I$ filters to calculate synthetic magnitudes and derive the magnitude differences, namely $\delta U$, $\delta B$, and $\delta I$, between the comparison and the reference spectrum. Isochrones 2--4 are derived by adding to the isochrone I1 the corresponding values of $\delta U$, $\delta B$, and $\delta I$. 
 
 In addition, we used two isochrones from \citet{dotter2008a} that differ from I1 because a variation in helium alone (I5) or in iron abundance alone (I6). These isochrones would mimic the subpopulations of 1G stars with different chemical composition \citep[e.g.\,][]{milone2015a, milone2022a, marino2019a, legnardi2022a}.

\begin{table}
  \caption{Chemical composition of the six isochrones shown in Figure~\ref{fig:ChMgbTEO}. All isochrones have ages of 13 Gyr and [$\alpha$/Fe]=0.4.}\label{tab:TEO}

\begin{tabular}{ c c c c c c}
\hline \hline
 Isochrone & Y & [C/Fe] & [N/Fe] & [O/Fe] & [Fe/H]  \\
\hline
 I1  & 0.246  &   0.00  &  0.00 &    0.40   & $-$1.50 \\
 I2  & 0.256  &$-$0.05  &  0.53 &    0.35   & $-$1.50 \\
 I3  & 0.276  &$-$0.10  &  0.93 &    0.20   & $-$1.50 \\
 I4  & 0.306  &$-$0.50  &  1.21 & $-$0.10   & $-$1.50 \\
 I5  & 0.266  &   0.00  &  0.00 &    0.40   & $-$1.50 \\
 I6  & 0.246  &   0.00  &  0.00 &    0.40   & $-$1.45 \\

     \hline\hline
\end{tabular}
  \label{tab:chimica}
 \end{table}
%% %%%%%%%%%%%%%%%%%%%%%%%%%%%%%%%%%%%%%%%%%%%%%%%%%%%%%%%%%%%%%%%%%%%%%%%%%%

\begin{figure*}
\begin{center}
	\includegraphics[height=8.0cm,trim={0cm 0.5cm 0cm 8cm}]{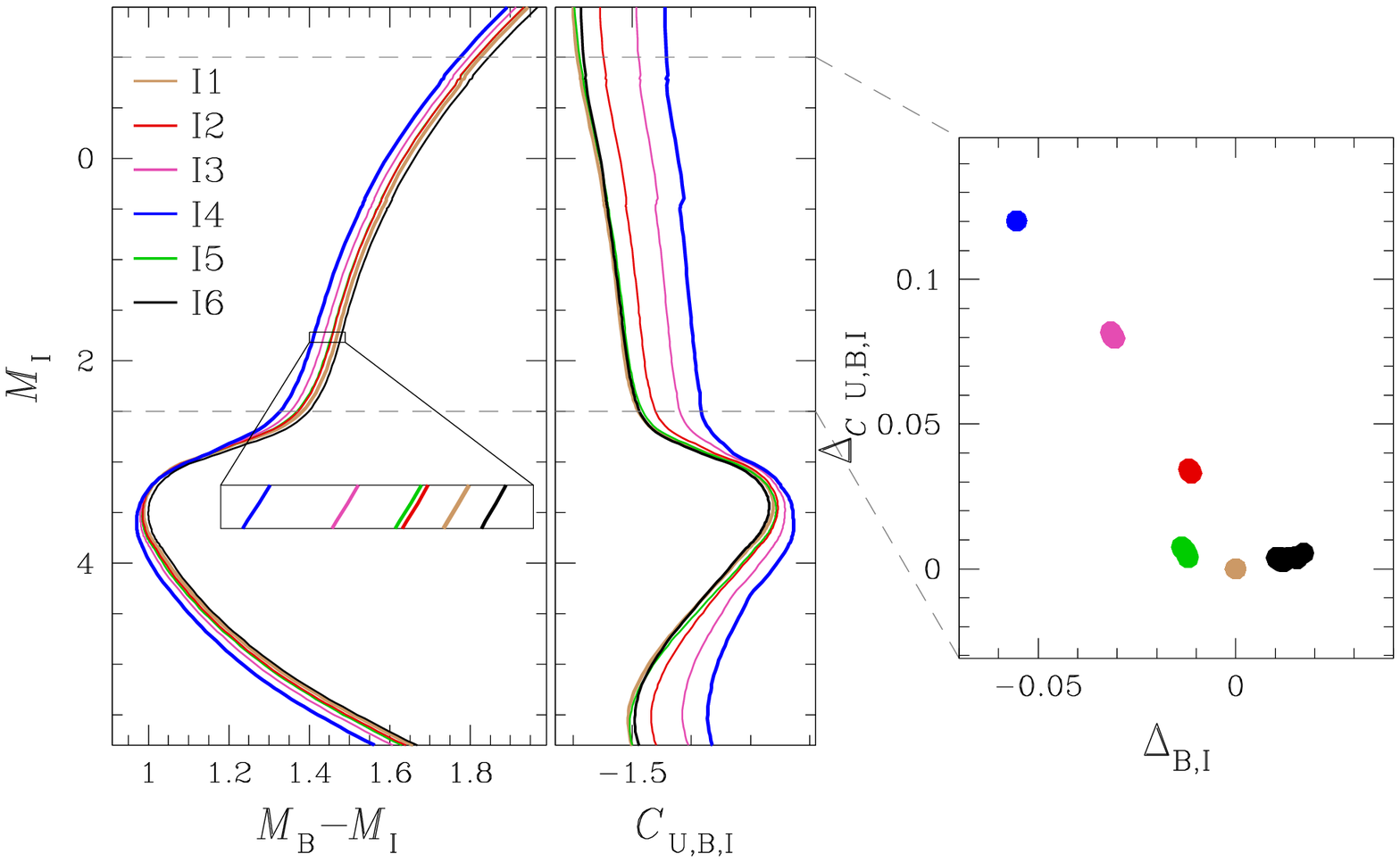}
    \caption{The six coloured lines plotted in the left and middle panel represent the isochrones I1--I6 (see Table\,\ref{tab:TEO} for details on their chemical compositions). The $\Delta_{C \rm U,B,I}$ vs.\,$\Delta_{\rm B,I}$ ChM of the RGB stars between the dashed horizontal lines is plotted in the right panel. }  
    \label{fig:ChMgbTEO}
\end{center}
\end{figure*}

The isochrones I1--I6 are plotted in the $M_{\rm I}$ vs.\,$M_{\rm B}-M_{\rm I}$ CMD and the $M_{\rm I}$ vs.\,$C_{\rm U,B,I}$ pseudo CMD of Figure\,\ref{fig:ChMgbTEO} (left and middle panel, respectively), where we also show the corresponding $\Delta_{C \rm U,B,I}$ vs.\,$\Delta_{\rm B,I}$ ChM for RGB stars. Clearly, the new ChM would allow us to separate 1G stars, which distribute around the origin of the reference frame, from 2G stars.

%%%%%%%%%%%%%%%%% BODY OF PAPER %%%%%%%%%%%%%%%%%%

Our analysis of synthetic spectra reveals that the $I$ vs.\,$B-I$ CMD is mainly affected by helium variations between 1G and 2G RGB stars. The $I$ vs.\,$C_{\rm U,B,I}$ diagram  of monometallic GCs is mostly sensitive to nitrogen variation among the different stellar populations.  
The ChM of Figure\,\ref{fig:ChMgbTEO} combines these two diagrams, thus providing more information on the chemical composition of multiple populations than each individual diagram. 

 Moreover, since the distinct stellar populations of most GCs have different values of $\Delta_{C \rm U,B,I}$ and $\Delta_{\rm B,I}$, the  ChM would provide a wider separation between 1G and 2G stars of most GCs rather then the $I$ vs.\,$B-I$ and the  $I$ vs.\,$C_{\rm U,B,I}$ diagrams, separately. In particular, it would allow us to detect  extended or multimodal sequences of 1G stars.

An additional advantage of the ChM is that the RGB stars of each simple stellar population would be distributed in a nearly point-like ares of the ChM. Hence, the distribution of stars in the ChM is poorly dependent on the evolutionary phase along the RGB.

\subsection{Comparison with the traditional ChM from {\it HST}}\label{sub:HST}
%To confirm whether our identification of multiple populations also corresponds to that of 1G and 2G identified from the 
 
 To further investigate the distribution of 1G and 2G stars along the ground-based ChM, we investigate the ChM stars for which both ground-based and {\it HST} photometry are available. It is widely accepted that 1G stars are clustered around the origin of the traditional ChM derived from {\it HST} photometry, while 2G stars are distributed towards larger and smaller values of $\Delta c_{\rm F275W,F336W,F814W}$ and $\Delta_{F275W,F814W}$, respectively \citep[][]{milone2015a}.
 
 As an example, probable 1G and 2G stars of NGC 288 are coloured red and blue, respectively, in the upper-left panel of Figure\,\ref{fig:tagging} where we show the $\Delta c_{\rm F275W,F336W,F814W}$ versus $\Delta_{F275W,F814W}$ from \citet{milone2017a}. 
 The upper-right panel of Figure\,\ref{fig:tagging} shows the $\Delta c_{\rm U,B,I}$ ChM of NGC\,288 derived in this paper (right).  The probable 1G and 2G stars identified in the upper-left panel and for which both ground-based and {\it HST} photometry is available are marked with large squares.
  
  Results from this comparison confirm the predictions from the simulated ChM of Section\,\ref{sub:teo}. 
  Indeed, the two groups of selected 1G and 2G stars populate the regions of the upper-right panel ChM with small and large values of $\Delta c_{\rm U,B,I}$ and $\Delta_{\rm B,I}$, respectively. 
Two stars that are classified as 2G in the {\it HST} ChM but exhibit low values of $\Delta c_{\rm U,B,I}$ together with one 1G star with large $\Delta c_{\rm U,B,I}$ values are possible exceptions. 

%  chromosome map, we have analysed stars that are included in the both of the chromosome maps. Open starred dots in left-panel of Figure \ref{fig:tagging} indicate stars that have also been analysed in the chromosome map $\Delta_{\rm C_{\rm F275W,F336W,F438W}}$ versus $\Delta_{\rm F275W,F814W}$, which are coloured in red or blue according to our identification of 1G and 2G as described above. These stars are marked on the chromosome map of NGC 288 derived by \citet{milone2017a} in the right panel using the same symbol, showing that 1G and 2G identified in this paper would be well matched those on the chromosome map built from {\it HST} photometry with three exceptions.  
% APM invertire ? (selezione 1G-2G in HST?)
% abbondanze medie di Na ed O

\subsection{Chemical tagging of stellar populations along the ChM of NGC\,288}\label{sub:spectroscopy}
In the past decades, the synergy of spectroscopy and photometry from both {\it HST} and ground-based facilities has provided major advances towards the understanding of the multi-population phenomenon. Early results based on wide-band ground-based photometry revealed that  Na-poor/O-rich and Na-rich/O-poor stars define different RGB sequences in the $B$ vs.\,$U-B$ CMDs of M\,4 and NGC\,6752 \citep[][]{marino2008a, milone2010a}. Moreover, 1G and 2G stars of 47\,Tucanae populate distinct sequences in the $B$ vs.\,$U-B+I$ diagram \citep[][]{milone2012b} and in the $V$ vs.\,$C_{\rm U,B,I}$ diagram \citep[][]{monelli2013}. In general, the sodium and oxygen abundances correlate and anticorrelate, respectively, with the $C_{\rm U,B,I}$ value, as shown by \citet[][]{monelli2013} for 15 GCs. Similar conclusion that the colour of RGB stars depends on light-element abundance is provided by pioneering works based on Str{\"o}emgren photometry \citep[][]{grundahl1998a, yong2008a, carretta2011a}.

Similarly, a lot of efforts have been dedicated to chemically characterize the stellar populations along the ChMs derived from both {\it HST} \citep[e.g.][]{milone2015a, carretta2018a, marino2019a}, and ground-based photometry \citep[][]{hartmann2022a}.

\begin{figure}
\begin{center}
	\includegraphics[height=9.0cm,trim={1cm 0cm 0cm 0cm}]{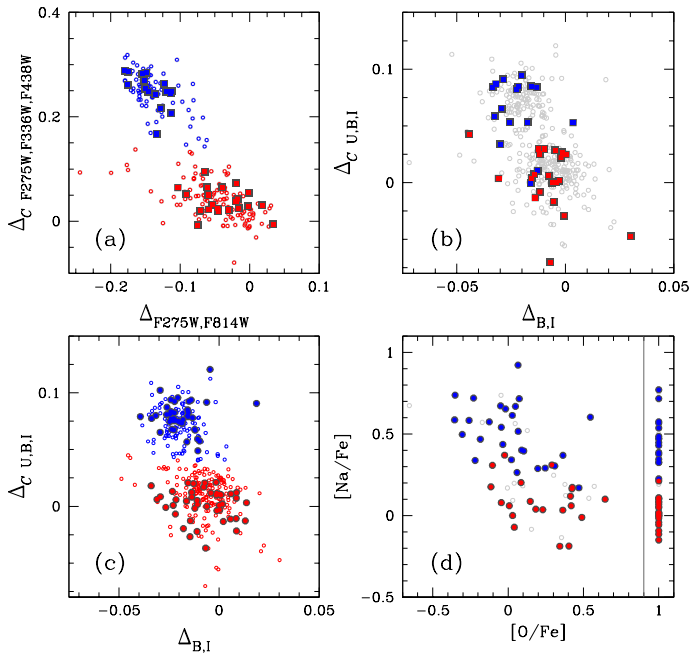}
    \caption{{\it Upper panels.} $\Delta c_{\rm F275W,F336W,F814W}$ versus $\Delta_{F275W,F814W}$ ChM of NGC\,288 from {\it HST} photometry \citep[][]{milone2017a}. The probable 1G and 2G stars selected by Milone and collaborators are coloured red and blue, respectively (upper-left). Reproduction of the $\Delta c_{\rm U,B,I}$ and $\Delta_{\rm B,I}$ ChM derived in Figure \ref{fig:procedureChM}.
    Large squares represent stars with both {\it HST} and ground-based photometry (upper-right).
    {\it Bottom panels.} The 1G and 2G stars identified in the $\Delta c_{\rm U,B,I}$ and $\Delta_{\rm B,I}$ plane are represented with red and blue circles, respectively, in the bottom-left ChM.
    Sodium versus oxygen abundances for RGB stars of NGC\,288 \citep{carretta2009a}.    %Left Panel shows the chromosome map of RGB stars in NGC 288, where we have coloured red and blue 1G and 2G stars, respectively. Filled red and blue triangles indicate 1G and 2G stars studied spectroscopically by \citet{carretta2009a}, and whose [Na/Fe] versus [O/Fe] anticorrelation is shown in middle panel using the same symbols. 
    Stars for which [O/Fe] estimates are not available are plotted on the right side of the vertical line in the panel. 
    Stars included in the ChM for which sodium abundances are available are represented with large dots.
    %Open red and blue starred symbols in the left panel are used for 1G and 2G studied also on the $\Delta c_{\rm F275W,F336W,F814W}$ versus $\Delta_{F275W,F814W}$ ChM derived by \citet{milone2017a}, which is illustrated in right panel with these stars using the same symbols.
    }
    \label{fig:tagging}
\end{center}
\end{figure}

In the bottom-left panel of Figure\,\ref{fig:tagging} we use the $\Delta c_{\rm U,B,I}$ and $\Delta_{\rm B,I}$ ChM to identify by eye the probable 1G and 2G stars of NGC\,288 (red and blue circles, respectively), based on their position on the ground-based ChM alone.
 
 To constrain their chemical composition, we use the chemical abundances inferred by \citet{carretta2009a} from high-resolution spectroscopy. Some results are illustrated in the bottom-right panel of  Figure\,\ref{fig:tagging}, where we plot [Na/Fe] vs.\,[O/Fe]. Stars for which both photometry and spectroscopy is available are plotted with large dots. Clearly, 1G stars are Na-poor and O-rich, whereas 2G stars are sodium enhanced and oxygen depleted. 
 The average abundances for the 1G stars and the 2G stars are listed in Table\,\ref{tab:NaOFe} and are in agreement %[Na/Fe]$_{\rm 1G}$= 0.05 $\pm$ 0.02, [Na/Fe]$_{\rm 2G}$= 0.47 $\pm$ 0.03, [O/Fe]$_{\rm 1G}$= 0.34 $\pm$ 0.04, [O/Fe]$_{\rm 2G}$= 0.06 $\pm$ 0.08, [Fe/H]$_{\rm 1G}$=$-$1.21 $\pm$ 0.01, and [Fe/H]$_{\rm 2G}$=$-$1.22 $\pm$ 0.01. 
 %This is in agreement 
  with the results from \citet{marino2019a}, where 1G and 2G stars have been identified on the traditional ChM from ({\it HST}). However, the average abundances inferred in this paper have smaller errors than those provided by Marino and collaborators, thanks to the large number of spectroscopic targets for which ground-based photometry is available.
  %photometry and combined with the spectroscopic data from \citet{carretta2009a}, [Na/Fe]$_{\rm 1G}$= 0.10 $\pm$ 0.10, [Na/Fe]$_{\rm 2G}$= 0.51 $\pm$ 0.14, [O/Fe]$_{\rm 1G}$= 0.25 $\pm$ 0.16, [O/Fe]$_{\rm 2G}$= 0.11 $\pm$ 0.26 [Fe/H]$_{\rm 1G}$=$-$1.21 $\pm$ 0.02, and [Fe/H]$_{\rm 2G}$=$-$1.23 $\pm$ 0.05. Note that, when combined with spectroscopic data, ground-based photometry provide a larger sample of stars available due to a wide field of view, providing smaller errors of all the abundances.  
 
 \begin{table}
  \caption{Average abundances of oxygen, sodium, and iron of 1G and 2G stars in NGC\,288 inferred in this paper and by \citet[][]{marino2019a}.}
%APMj10: Uncertainties from M19 seems very big. Please check that we are really using the errors and NOT the r.m.s (sigma).  
\begin{tabular}{ c c c c }
\hline \hline
     & [O/Fe] &  [Na/Fe] & [Fe/H]  \\
     \hline
     &        &  This paper         &          \\
1G   & 0.34$\pm$0.04  &  0.05$\pm$0.02      &  $-$1.21 $\pm$ 0.01        \\
2G   & 0.06$\pm$0.08  &  0.47$\pm$0.03      &  $-$1.22 $\pm$ 0.01        \\
\hline
     &        &  \citet[][]{marino2019a}        &          \\
1G   &  0.25$\pm$0.09      &   0.10$\pm$0.04      &  $-$1.21$\pm$0.01        \\
2G   &  0.11$\pm$0.13      &   0.51$\pm$0.04      &  $-$1.23$\pm$0.02        \\
     \hline\hline
\end{tabular}
  \label{tab:NaOFe}
 \end{table}

\begin{figure}
\begin{center}
	\includegraphics[height=9.0cm,trim={0cm 0cm 0cm 0cm}]{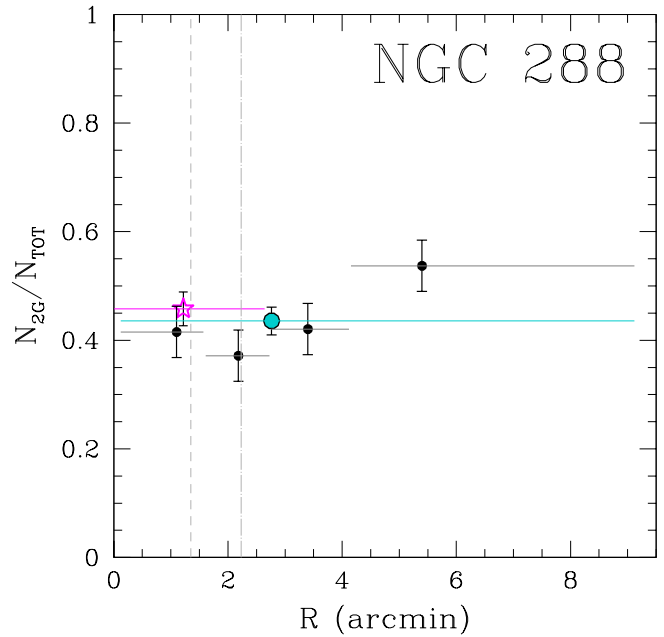}
    \caption{ Fraction of 2G stars of NGC\,288 against the radial distance from the cluster center. Black dots are derived in this paper, whereas the magenta starred symbol indicates the fraction of 2G stars measured by \citet{milone2017a} from {\it HST} photometry. The cyan dot marks the average fraction of 2G stars from ground-based photometry. The horizontal segments mark the radial interval corresponding to each point. Vertical dotted and dashed-dotted lines indicate the core and half-light radius \citep[from the 2010 version of the][catalog]{harris1996a}. }
    \label{fig:RD}
    % HST point: ---> average Milone+2017 and Piotto+2013.
    % HST point: ---> centered on the average radius of MS-RGB stars.
\end{center}
\end{figure}

\begin{figure*}
\begin{center}
	\includegraphics[height=18.5cm,trim={0cm 0cm 0cm 0cm}]{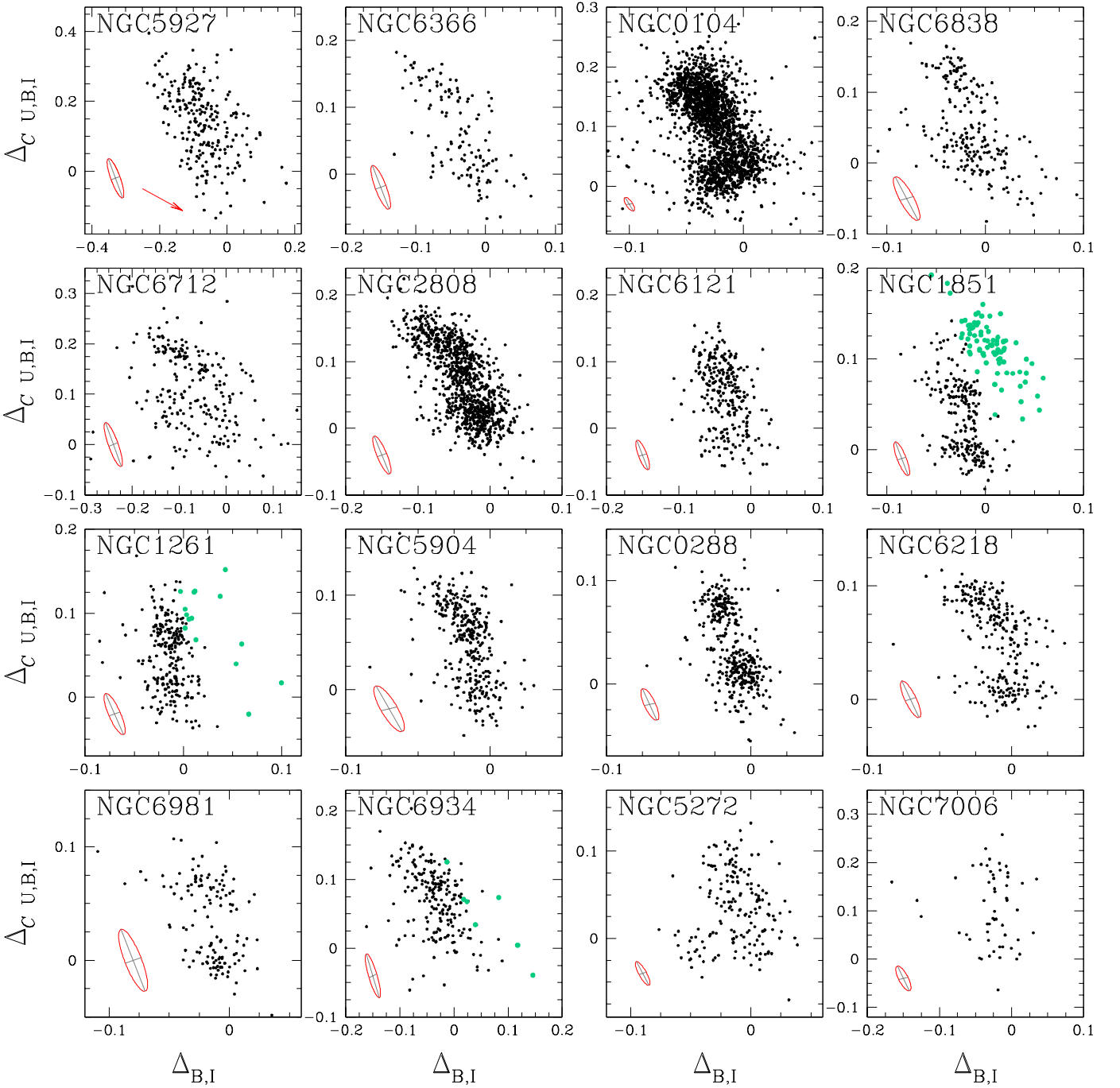}
    \caption{$\Delta c_{\rm U,B,I}$ versus $\Delta_{\rm B,I}$ diagrams (or chromosome maps) for RGB stars in the GCs NGC\,5927, NGC\,6366, NGC\,104, NGC\,6838, NGC\,6712, NGC\,2808, NGC\,6121, NGC\,1851, NGC\,1261, NGC\,5904, NGC\,288, NGC\,6218, NGC\,6981, NGC\,6934, NGC\,5272, and NGC\,7006. The clusters are sorted according by metallicity, from the most metal-rich to the most metal-poor. Aqua symbols mark red-RGB stars of Type\,II GCs. The arrow n the top-right panel indicates the reddening vector and correspond to a reddening variation $\Delta$E($B-V$) = 0.05 mag}.
    \label{fig:chms1}
\end{center}
\end{figure*}

\begin{figure*}
\begin{center}
	\includegraphics[height=18.5cm,trim={0cm 0cm 0cm 0cm}]{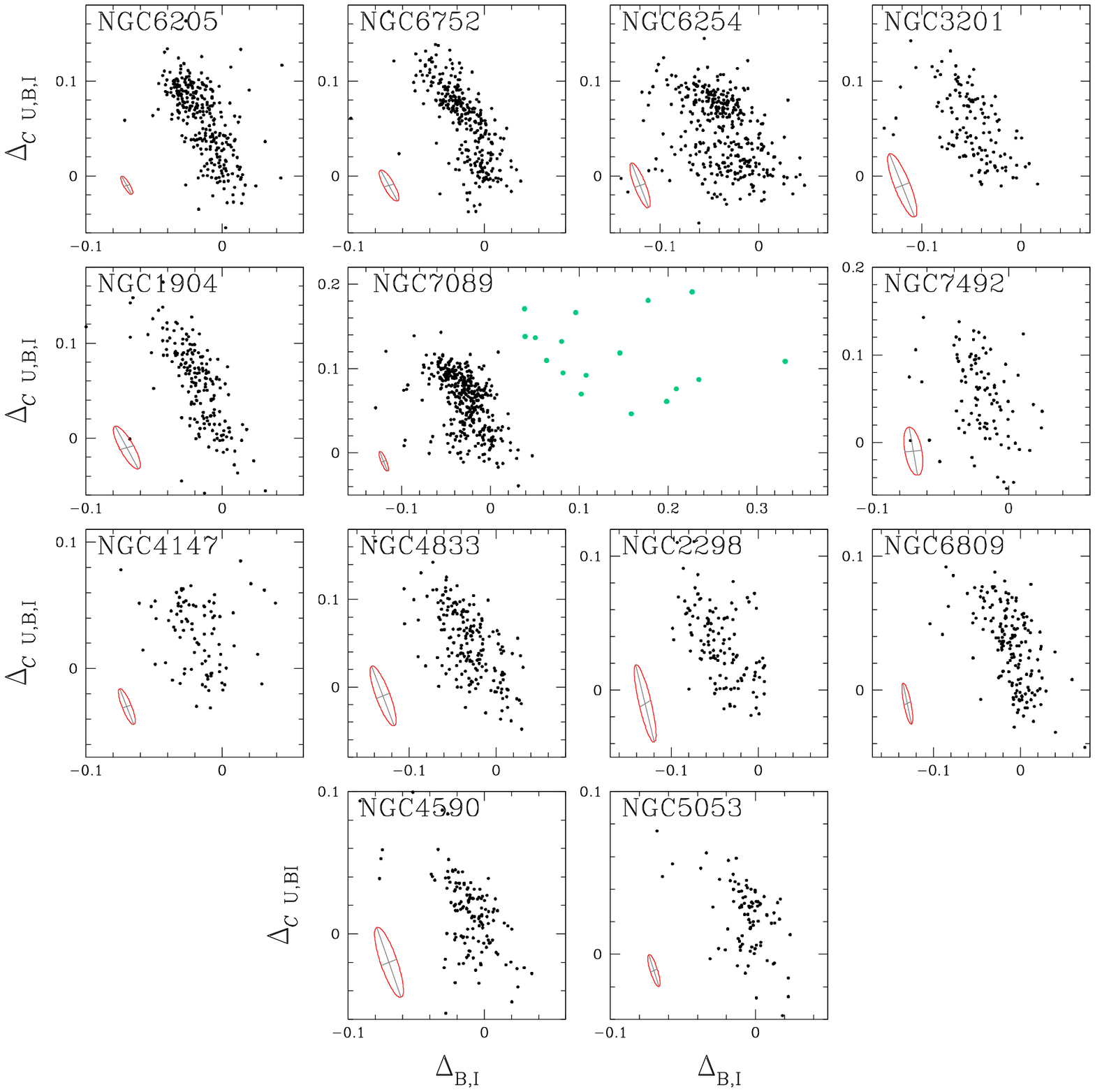}
    \caption{As in Figure \ref{fig:chms1}, but for NGC\,6205, NGC\,6752, NGC\,6252, NGC\,3201, NGC\,1904, NGC\,7089, NGC\,7492, NGC\,4147, NGC\,4833, NGC\,2298, NGC\,6809, NGC\,4590, and NGC\,5053.}
    \label{fig:chms2}
\end{center}
\end{figure*}

\subsection{The radial distribution of multiple populations in NGC\,288}\label{sub:RD} 
Most scenarios for the formation of multiple populations predict that 2G stars form in the cluster center \citep[e.g.][and references therein]{dercole2008a, denissenkov2014a, dantona2016a, gieles2018a, renzini2015a}. Due to dynamic evolution, some GCs can retain some memory of the initial distribution of  1G and 2G stars, while the multiple populations of other clusters can fully mixed \citep[e.g.][]{vesperini2013a, dalessandro2019a}. Hence, the present-day spatial distribution of multiple populations in GCs would provide   valuable tests for the formation scenarios. Work based on both spectroscopy and photometry show that 2G stars of various GCs (e.g.\,47\,Tucanae, NGC\,2808, M\,3, NGC\,5927 among the others)  \citep[e.g.][]{cordero2014a, milone2012a, vanderbeke2015a, lee2017a, dondoglio2021a} are more centrally concentrated than the 1G, whereas multiple populations of other clusters (e.g.\,NGC\,6752, NGC\,6362, M\,5, NGC\,6366, NGC\,6838) share the same radial distribution \citep[e.g.][]{vanderbeke2015a, lee2017a, milone2019a, dalessandro2018a, cordoni2020a, dondoglio2021a}. Some authors suggest that a few clusters can exhibit reverse radial distribution in contrast with  theoretic predictions \citep[see][for the case of NGC\,3201]{hartmann2022a}.

%APMjul10: I think that the old text is not correct. Either we use a bi-Gaussian fit and compare the area of the 2G Gaussian and the total area. Or (I believe) we used the method by Milone+2017, which is based on a single Gaussian. If I remember, we did NOR rotate di ChM of NGC288 but we did for 47 Tuc and other clusters. I would then keep a generic text.
To derive the population ratios in NGC\,288 we used the method by \citet[][see their Figure\,8]{milone2017a}. We first rotated the ChM in a reference frame, $\Delta_{2}$ vs.\,$\Delta_{1}$, where the abscissa follows the direction of 1G stars. Then, we derived the histogram distribution of the ordinate $\Delta_{2}$ and fitted the histogram of candidate 1G stars with a Gaussian function by means of least squares. The fraction of 1G stars is derived as the ratio of the area of the Gaussian over that of the whole histogram.
 %least-squares method to fit the distribution of the $\Delta c_{\rm U,B,I}$ pseudo-colour with a function made by the sum of two Gaussian functions. The fraction of 2G stars is provided by the ratio between the area of the Gaussian function that provides the best match with 1G stars and the total area of the best-fit bi-Gaussian function.  
%%%

We find that 2G stars comprise 44$\pm$2\% of the total number of cluster stars, in the entire region within 10 arcmin from the cluster center. This value is consistent with previous estimates based on MS \citep[46$\pm$3\%,][]{piotto2013a} and RGB stars \citep[46$\pm$3\%,][]{milone2017a} in the central field of view observed with {\it HST} (radius smaller than 2.7 arcmin, but covering the cluster half-light radius $r_{\rm h}$=2.23 arcmin).

The wide field of analyzed in this paper provides the opportunity to investigate the radial distribution of the population ratio. To do this, we divided the field of view into four radial intervals including the same numbers of stars in the ChM and calculate the fraction of 2G stars in each interval.
Results are illustrated in Figure\,\ref{fig:RD} where we plot the fraction of 2G stars in NGC\,288 as a function of the radial distance from the cluster center. Results are consistent with a constant fraction of 2G stars within the analyzed radial interval. We note an hint for higher fraction of 2G stars outside five arcmin from the cluster center but this results is significant at $\sim$1.5$\sigma$ level only. 
% APMjul10: I would drop the following connection with kinematics.
%           Indeed: i) Our result is NOT a statistically significant (the disagrament is 1.5 sigma only). ii) Our results is not confirmed by HST data of VLM stars (Emanuele) and bright MS stars (from myself).
% iii) Results on velocity profile are not significant as well as clearly pointed out by Laura Watkins et al. 
%It might be related to the observed velocity dispersion radial profiles of NGC 288 obtained with {\it HST}-based proper motion that have a similar radial trend with the fraction of 2G stars  \citep[][]{libralato2022a,wak2015}.

\section{The atlas of RGB chromosome maps}\label{sec:atlas}
Figures\,\ref{fig:chms1} and \ref{fig:chms2} illustrate a collection of ChMs for 29 GCs, including four Type-II GCs, where it is possible to disentangle the bulk of 1G and 2G stars \footnote{As shown in Figure\,9, the pseudo-colour separation between 1G and 2G stars in the ChM depends on the relative content of some light elements (mostly He, N, C). Moreover, for a fixed variation of Y, [N/Fe] and [C/Fe],  the magnitude and colour differences between 1G and 2G stars are larger in metal-rich GCs rather than in metal poor ones \citep[][]{milone2018a}.
Hence, we would expect a small separation between 1G and 2G stars in the ChM of metal poor clusters. 
i) Most clusters that we excluded from the ChM analysis are indeed metal-poor GCs. Eight out fourteen clusters, namely NGC\,4372, NGC\,5024, NGC\,5634, NGC\,6101, NGC\,6341, NGC\,7078, NGC\,7099, and Terzan\,8, have very low iron abundances of [Fe/H]$<-$1.95, \citep[2010 version of the][catalog]{harris1996a}.
ii) Four targets (IC\,4499, NGC\,5286, NGC\,5986, and NGC\,6656) are metal-poor GCs with  $-$1.70$<$[Fe/H]$<-$1.53 and large reddening (E(B$-$V)=0.23$-$0.34 mag). In addition to the relatively small separation between 1G and 2G stars in the ChM that we would expect in metal-poor GCs, we note that their magnitudes (in particular in the U band) exhibit larger errors when compared with most studied GCs.
iii) On the other hand, NGC\,6760 and Palomar\,11 are metal-rich ([Fe/H]=$-$0.40), but are highly obscured by interstellar clouds (E(B$-$V)=0.77 and 0.35 mag, respectively). Hence, have poorer photometry than the bulk of analyzed GCs.}. 

 The clusters are sorted in order of decreasing metallicity, from most metal-rich (NGC\,5927, [Fe/H]=$-$0.49) to most metal-poor (NGC\,5053, [Fe/H]=$-$2.27). 
 Due to crowding, the photometric quality of the central regions can be too poor to derive accurate ChMs. Hence, for some clusters we derive the ChM by using stars that are located outside a certain radius that we fixed by eye. Moreover, we excluded the few RGB stars outside the region used to infer the differential reddening.
 We provide the radius range of stars used to build the ChM of each cluster, $R_{\rm ChM}$, in Table\,\ref{tab:DR}. 
 
 A visual inspection at the ChMs reveals a great deal of variety, in close analogy with what is observed in the $\Delta c_{\rm F275W,F336W,F814W}$ versus $\Delta_{F275W,F814W}$ ChMs \citep[][]{milone2017a}. Specifically:
 
 \begin{itemize}
     \item  The $\Delta c_{\rm U,B,I}$ and $\Delta_{B,I}$ extensions of the ChMs change from one cluster to another. Typically, metal-rich GCs exhibit more-extended ChMs than metal poor GCs, but there are clusters with similar metallicities (e.g.\,M\,4 and NGC\,2808) but different ChM shapes.
     
     \item
     The fractions of probable 1G stars (i.e.\, stars located around the origin of the ChM) changes from one cluster to another. As an example, the majority of stars in NGC\,6366 and NGC\,6838 belong to the 1G, while NGC\,2808 and NGC\,104 are dominated by 2G stars.
     The distribution of stars along the 2G is typically continuous. However some hints of double or triple populations of 2G stars are present in NGC\,6205, NGC\,6752, and NGC\,2808. 
 
      \item
      The red-RGB stars of Type\,II GCs that we selected in Figure\,\ref{fig:cmddr} (aqua dots in Figures\,\ref{fig:chms1} and \ref{fig:chms2}) define distinct ChM sequences with redder  $\Delta_{\rm B,I}$ values than the bulk of blue-RGB stars.
 
 \item The relative $\Delta_{\rm B,I}$ extensions of 1G and 2G stars differs from one cluster to another.
 In several GCs (e.g.\,NGC\,6218 and NGC\,2808) the 2G sequence spans a wider range of $\Delta_{\rm B,I}$ pseudo colours than 1G stars, whereas the 1G and 2G sequences of other clusters exhibit similar $\Delta_{\rm B,I}$ extensions (e.g.\,NGC\,288 and NGC\,1261). 
 In NGC\,6366 and  NGC\,6838 the 1G sequence is more-extended than the 2G along the $\Delta_{\rm B,I}$ axis, while the 1G stars of NGC\,6712 and NGC\,6254 exhibit wider $\Delta c_{\rm U,B,I}$ than the 2G. Since 1G and 2G stars have similar observational errors, this fact proves that 1G stars are not chemically homogeneous \citep[][]{milone2017a, milone2018a, marino2019b, legnardi2022a}.
 
 \item Intriguingly, the slope of the 1G sequence in the ChMs of GCs with extended 1G sequences seems to vary from cluster to cluster. In most GCs with extended 1G sequences (e.g.\,NGC\,6366 and NGC\,6838) the 1G stars with $\Delta_{\rm B,I} \sim 0$ have smaller values of $\Delta c_{\rm U,B,I}$  than 1G stars with negative $\Delta_{\rm B,I}$ values. A similar trend is observed in the traditional ChMs of all clusters \citep[][]{milone2018a}. 
  Noticeably, the trend is reversed in NGC\,5927 and NGC\,104.
 
  \end{itemize}

\begin{figure}
\begin{center}
	\includegraphics[height=9.0cm,trim={0cm 0cm 0cm 0cm}]{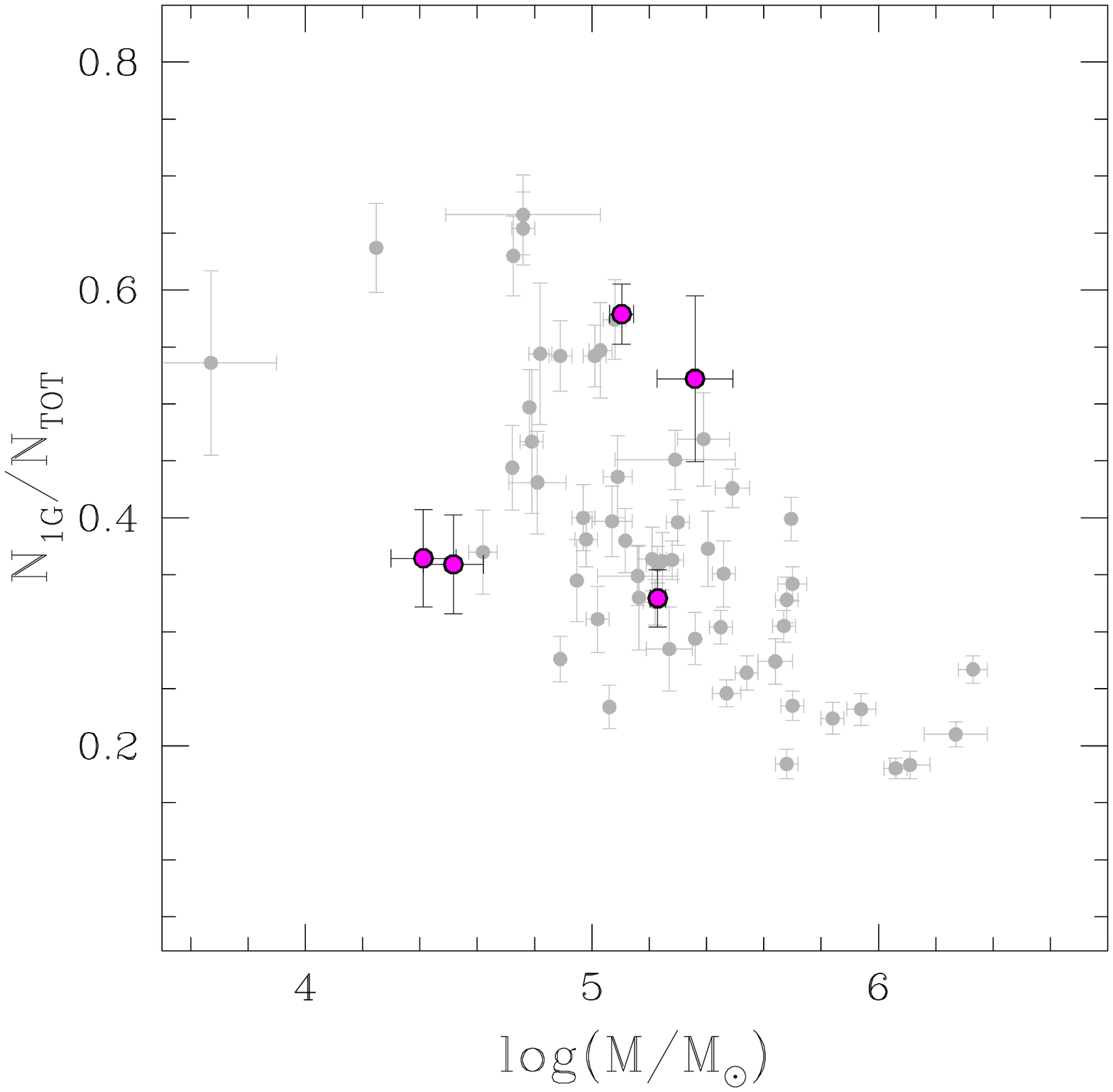}
    \caption{Fraction of 1G stars against the present-day cluster mass for Galactic GCs. Gray points indicate literature results \citep[][]{milone2017a, dondoglio2021a}, whereas magenta dots mark the clusters without previous studies on the ChM. Cluster masses are from \citet[][]{baumgardt2018a}.}
    % APM. I would remove the points for NGC288 and 47Tuc.
    % APM. I would use \odot instead of msun on the xlabel label. 
    \label{fig:fraction}
\end{center}
\end{figure}

 Our sample includes five GCs, namely NGC\,1904, NGC\,4147, NGC\,6712, NGC\,7006 and NGC\,7492, without previous determinations of the ChM. We estimate their fractions of 1G stars, which are provided in Table \ref{tab:tab2}.
 %, which range between 33$\pm$3 \% to  58$\pm$3. 
 These clusters are highlighted with magenta dots in Figure\,\ref{fig:fraction}, where we plot the fractions of 1G stars \citep[from][]{milone2017a, dondoglio2021a} vs.\, the GC mass \citep[][]{baumgardt2018a}. Clearly, the newly studied clusters follow the same relation between the fraction of 1G stars and cluster mass as the bulk of Galactic GCs \citep[][]{milone2017a, milone2020a}.

 %, contradicting with a case of NGC\,288. 
\begin{figure*}
\begin{center}
    % APM: I would suggest some slight changes to this Figure.
    %      -- remove panel a and plot the three panels on the same line.
    %      -- panels b and c could have the same scale on the y axis.
    %      -- different symbols or colours for the bright AGB stars that are enot included in the ChM
	\includegraphics[height=6.0cm,trim={0cm 13.5cm 0cm 0cm}]{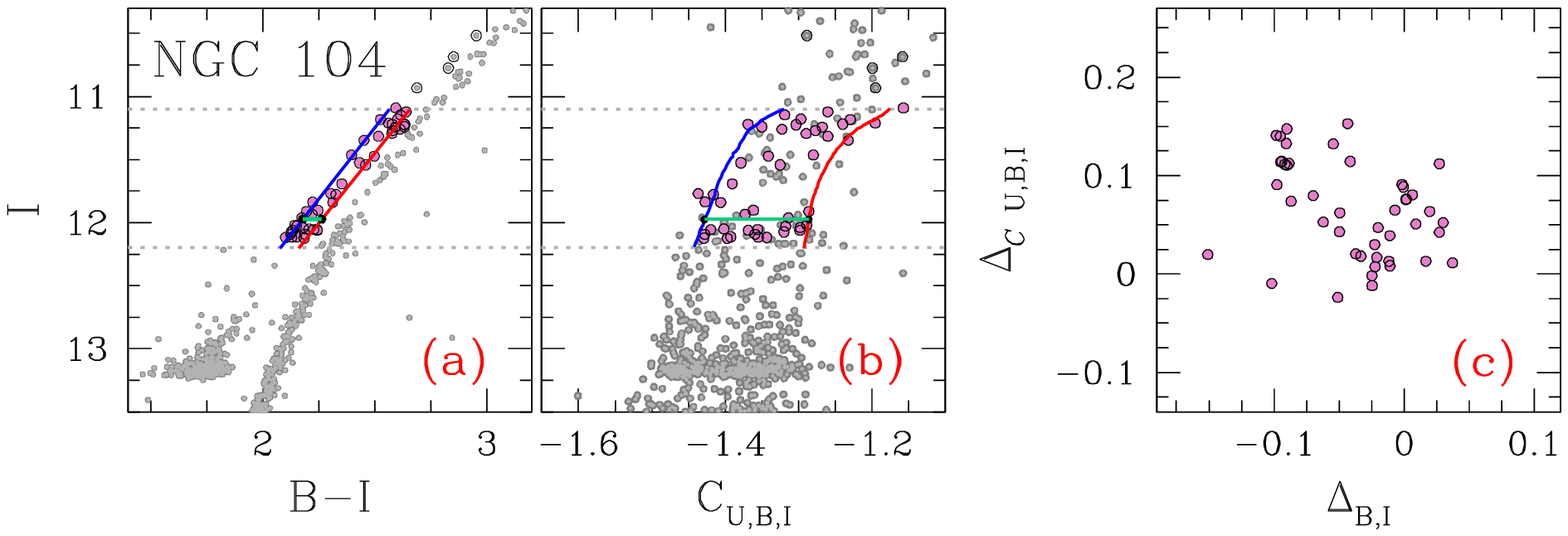}
    \caption{This figure illustrates the procedure to derive the ChM of AGB stars in NGC\,104 (47\,Tucanae).  Panels a and b represent the $I$ vs.\,$B-I$ and $I$ vs.\,$C_{\rm U,B,I}$ diagrams, respectively. The red and blue lines mark the boundaries of the AGB sequence, while the aqua segments indicate the width of the AGB sequence. AGB stars are marked with dots. Only AGB stars between the two horizontal (pink dots) lines are used to derive the ChM plotted in panel c. }
    \label{fig:agb}
\end{center}
\end{figure*}

\section{The chromosome map of AGB stars}\label{sec:AGB}
The investigation of multiple populations along the AGB represents crucial test for stellar evolution. Stellar evolution models predict that most GC stars would evolve into the AGB phase, possibly, with the remarkable exception of 2G stars stars with extreme helium contents \citep[e.g.][]{kippenhahn1990a, landsman1996a, chantereau2016a}. 
 This conclusion is challenged by spectroscopic work on multiple population, which reveal that some 2G stars with moderate helium enhancement skip the AGB phase \citep[see e.g.][]{campbell2013a, johnson2012a, joh2015, marino2017a}.

Recent works, based on {\it HST} photometry, have shown that ChMs are efficient tools to identify and characterize multiple populations along the AGB of GCs \citep[][]{marino2017a, lagioia2021a}.
In the following, we present the method to derive ChM  of AGB stars from $U, B, I$ ground-based photometry.

Due to the small number of AGB stars in NGC\,288,  we use NGC\,104 as a test case. The procedure is illustrated in Figure\,\ref{fig:agb} and is similar to the method that we used for RGB stars.
In a nutshell, we first selected by eye a sample of bona-fide AGB stars, based on their position in the $I$ vs.\,$B-I$ CMD (Figure\,\ref{fig:agb}a). The ChM is derived from the $I$ vs.\,$B-I$ and the $I$ vs.\,$C_{\rm U,B,I}$ diagrams shown in panels a and b of Figure\,\ref{fig:agb}. 
 The red and blue boundaries of the AGB are derived by hand and are used to derive the $\Delta_{\rm B,I}$ and $\Delta c_{\rm{U,B,I}}$ pseudo colours by using Equations\ref{eq:quadratic} and \ref{eq:quadratic2}. The values of $W_{\rm B,I}$ and $W_{C \rm U,B,I}$ correspond to the $B-I$ and $C_{\rm U,B,I}$ separations between the fiducial lines, calculated 5 $I$ mag above the main-sequence turnoff.

The resulting ChM for AGB stars in NGC\,104 is plotted in panel c of  Figure\,\ref{fig:agb}. Clearly, AGB stars comprise a group 1G stars that are distributed around the origin of the ChM and a group of 2G stars with large values of $\Delta c_{\rm{U,B,I}}$ and  $\Delta_{\rm B,I}$.
Noticeably, the fraction of 1G stars along the AGB is 58$\pm$5 \% and is significantly larger than that derived from the ChM of RGB stars (40$\pm$1 \%).
This is in agreement with the spectroscopic study of NGC\,104 from \citet{joh2015}, where the fraction of Na-poor and Na-rich stars changes from 45:50 on the RGB to 63:37 on the AGB. Although this phenomenon still misses a definitive explanation, our results support the idea that 2G stars are affected by larger amounts of RGB mass loss than 1G stars \citep[][]{campbell2013a, joh2015, tailo2015a, tailo2020a}.  
It is also interesting to notice that the fractions of 1G inferred from wide-field photometry is much larger than that obtained from the traditional ChM in the  central region \citep[18$\pm$1 \% from][]{milone2017a}. This results is quite expected given that 2G stars of NGC\,104 are more centrally concentrated than the 1G ones \citep[][]{milone2012b,cordero2014a, dondoglio2021a}. When we estimate the population ratio from the ground-based ChM of the stars in the {\it HST} field of view, we find a fraction of 1G stars of 21$\pm$3 \%, in agreement with the values provided by the papers quoted above, in the same region.

\section{Summary and conclusions}\label{sec:conclusion}
By far, the photometric catalogs provided by Peter Stetson  have provided state-of-the-art photometry of GC stars from ground-based facilities \citep[][]{stetson1988a, stetson2005, stetson2019}.
We have analyzed photometry and astrometry of 43 GCs in the $U, B, V,$ and $I$ bands by \citet[][]{stetson2019} %to  identify and investigate multiple stellar populations over a wide field of view.  Photometry and astrometry by Stetson and collaborators have been 
 and combined them with stellar proper motions and parallaxes from Gaia eDR3 \citep{gaia2021a}. 
 
 We identified bona-fide cluster members and estimated the amount of differential reddening in the field of view of each cluster and found that the maximum reddening variation ranges from $\Delta$E(B$-$V)$\sim 0.19$ mag to less than 0.01 mag. The amount of differential reddening correlates with the average reddening in the direction of the cluster. Nevertheless, clusters with similar values of E(B$-$V) can exhibit different amounts of differential reddening. We corrected the photometry of 18 GCs with significant variation of reddening and publicly release the differential-reddening catalogs.
 
 Clearly, high-resolution reddening maps for GC cluster members have various astrophysical applications. As an example, applying the InfraRed Flux Method to better determine the effective-temperature scale of the RGB will have several implications for the study of stellar populations and stellar modelling \citep[e.g.][]{salaris2018a, casagrande2021a}. In this work, we focus on the multiple-population phenomenon.
 
To investigate multiple populations from ground-based photometry, we started using NGC\,288 as a test case. The reason for this choice is that NGC\,288 is a quite simple cluster in the context of multiple populations, where 1G and 2G stars exhibit distinct sequences in the photometric diagrams that are commonly used to investigate the multiple populations \citep[][]{piotto2013a, monelli2013, milone2017a, cordoni2020a}.
 The main results on NGC\,288 can be summarized as follows.

\begin{itemize}
    \item Based on the $I$ vs.\,$B-I$ CMD and the $I$ vs.\,$C_{\rm U,B,I}$ pseudo CMD of NGC\,288, we built the $\Delta c_{\rm U,B,I}$ vs.\,$\Delta_{\rm B,I}$ ChM for RGB stars. 
 The NGC\,288 stars distribute into two main blobs in the ChM. We matched the catalog from \citet[][]{stetson2019} and from {\it HST} photometry \citep[][]{milone2017a} to identify stars with both {\it HST} and ground-based photometry.
 We find that the 1G stars defined by \citet[][]{milone2017a} distribute around the origin of the  $\Delta c_{\rm U,B,I}$ vs.\,$\Delta_{\rm B,I}$ ChM, while 2G stars exhibit large values of  $\Delta c_{\rm U,B,I}$ and $\Delta_{\rm B,I}$. This is the first time that a ChM is derived from ground-based Johnson-Cousin photometry (see also the work by \citet{hartmann2022a} who derived the ChM from ground-based photometry, by using  the $U$ band and the narrow-band filters centered around 3780\AA\, and 3950\AA.).
  
  \item We also show that the $\Delta_{\rm U,B}$ vs.\,$\Delta_{\rm B,I}$ ChM, which is derived by the $I$ vs.\,$B-I$ and $I$ vs.\,$U-B$ CMDs, provides a clear separation between 1G and 2G stars.
  
  \item We used the values of [O/Fe], [Na/Fe], and [Fe/H] inferred by \citet[][]{carretta2009a} from high-resolution spectroscopy to infer the  chemical composition of the stellar populations identified on the ChM. We find that 1G stars and 2G stars have the same iron abundance ($\Delta$[Fe/H]$_{\rm 2G-1G}$=$-0.01 \pm 0.01$). 2G stars are sodium enhanced ($\Delta$[Na/Fe]$_{\rm 2G-1G}$=$0.42 \pm 0.04$) and depleted oxygen depleted ($\Delta$[O/Fe]$_{\rm 2G-1G}$=$-0.28 \pm 0.09$) with respect to the 1G.
  This result is consistent with previous findings by \citet[][]{marino2019a} based on the {\it HST} ChM and the elemental abundances from \citet[][]{carretta2009a}. However, the largest field of view covered by the ground-based photometry used in this paper provides improved results. Indeed, we used larger samples of 1G and 2G stars for which spectroscopic estimates of chemical composition are available.
  
  \item We studied the radial distribution of stellar populations of NGC 288 and find that the fraction of 1G stars is consistent with a flat distribution.  
  Similarly to NGC\,288, the 1G and 2G stars of various GCs \citep[e.g., NGC\,6752, NGC\,6362, M5, NGC\,6366, NGC\,6838 among the others][]{lee2017a, milone2019a, dalessandro2018a, dondoglio2021a} share the same radial distribution.
  These observations are consistent with a scenario where both populations born with the same radial distributions. However, a lack of radial gradient of the population ratio is also expected in the scenarios where 2G stars form in the innermost cluster regions. These scenarios predict that the multiple populations of some clusters are fully mixed due to dynamical evolution \citep{vesperini2013a}. 
  
  Results on NGC\,288 are consistent with the scenario by \citet[][]{hen2015}.
  Indeed, according to these authors (see their Figure 16), NGC 288 is on the edge between the evaporation and expansion dominated regions of the plot. 
  %The values of the mass and Galactocentric radius used to locate the cluster on the graph are taken from Baumgardt et al. 2018. 
  A cluster that is in the expansion-dominated phase of its evolution is believed to be not fully mixed. On the contrary, clusters in the evaporation-dominated phase are believed to be completely relaxed and to have their populations mixed, after losing a large fraction of their initial mass. Therefore, NGC 288 might have reached a significant mixing between its stellar populations, explaining the observed flat radial distribution of the 1G fraction. NGC\,104, on the other hand, is in the expansion-dominated phase of its evolution, therefore it still shows spatial and kinematic differences between the two populations, as shown in the radial distribution of NGC\,104 \citep[see Figure 19 of][]{dondoglio2021a}. This is also consistent with our finding that the fractions of 1G inferred from wide-field photometry is much larger than that obtained from the traditional ChM in the central region of NGC\,104.
 
\end{itemize}

  Driven by the results on NGC\,288, we %built the ChMs for all clusters analyzed by \citet[][]{stetson2019}. In this paper we focused on 
   present the ChMs of 29 GCs, where the bulk of 1G and 2G stars can be distinguished in the ChM. Our sample comprises five Type\,I GCs, namely NGC\,1904, NGC\,4147, NGC\,6712, NGC\,7006 and Palomar\,11, without previous determinations of ChMs.  For these clusters, we calculated the fraction of 1G stars and find that they follow the same relation with cluster mass as the other studied Galactic GCs.
  %Early analysis of the ChMs reveals the following multiple-population properties:
  % \begin{itemize}
      %\item  
      
      The atlas of 29 ChMs reveals that the extensions of the 1G and 2G sequences as well as the relative numbers of 1G and 2G stars change from one cluster to another.  We thus confirm that multiple populations exhibit a large degree of variety \citep[see also][]{renzini2015a, milone2017a, milone2020a}. Moreover, the ground-based ChMs allow to disentangle between Type\,I and Type\,II GCs. Indeed, the latter exhibit a ChM sequence with large $\Delta_{\rm B,I}$ values that run on the right side of the main ChM.

      We find that the 1G sequences of several GCs (e.g.\,NGC\,5927, NGC\,6366, NGC\,104, NGC\,6838, NGC\,6712, NGC\,5272, NGC\,6254, NGC\,3201, and NGC\,4833) are clearly elongated in the $\Delta c_{\rm U,B,I}$ vs.\,$\Delta_{\rm B,I}$  ChM, whereas the other GCs show less-extended 1G sequences along the abscissa of the ChM. We thus confirm the recent discoveries of extended 1G sequences in the ChM and corroborate the conclusion that the pristine material from which GC formed was not chemically homogeneous \citep[][]{milone2015a, milone2017a, cabreraziri2019a, legnardi2022a}. 
      
            Based on high-resolution spectroscopy and {\it HST} multi-band photometry, the presence of star-to-star metallicity variations seems the most plausible explanation for the 1G extended sequence \citep[][]{marino2019a, kamann2020a, legnardi2022a}. Helium variations and stellar binarity have been considered as alternative solutions, but they do not seem to fully reproduce the observations \citep[][]{milone2018a, marino2019a, kamann2020a}.
            
          The ChMs provided in this paper provide further constraints of the chemical composition of 1G stars. 
          We discovered that the 1G sequence of some clusters like NGC\,6366 and NGC\,6838 is thicker than the 2G sequence along the $\Delta c_{\rm U,B,I}$ direction.   Moreover, the slope of the 1G sequence of NGC\,104 and NGC\,5927  have positive slopes in the  $\Delta c_{\rm U,B,I}$ vs.\,$\Delta_{\rm B,I}$, in contrast with what is observed in the remaining GCs with extended 1G.
          
           In summary, results on NGC\,288 and on the other studied clusters demonstrate that the $\Delta c_{\rm U,B,I}$ vs.\,$\Delta_{\rm B,I}$  ChM is an efficient tool to identify and characterize multiple populations in GCs from ground-based  photometry. %As an example it allows us to investigate the spatial distributions of the stellar populations. Moreover, the synergy between the ChM and the elemental abundances inferred from spectroscopy makes it possible to derive the chemical composition of 1G and 2G stars. 
            The ChMs that we have introduced can be obtained from photometry from wide-field facilities. This fact allows us to overcome the main limitations on the study of multiple-populations that were   associated with the small field of view of the {\it HST} cameras.

\clearpage

 %\begin{landscape}
 \begin{table*}

\setlength{\tabcolsep}{15pt}
  \caption{This table lists for each cluster the values of the 68$^{\rm th}$ percentile, $\sigma_{\Delta E(B-V)}$, the difference between the 98$^{\rm th}$ and the 2$^{\rm nd}$ percentile of differential-reddening distributions, and the value of 68$^{\rm th}$ percentile of $\Delta E(B-V)_{\rm B-I}$ - $\Delta E(B-V)_{\rm U-V}$. The maximum radial distance used to derive differential reddening and the radius range used to build the ChM from cluster center of the region, $R_{\rm max}$ and R$_{\rm ChM}$, are also provided. 
  }
  \label{tab:DR}
  \begin{tabular}{cccccc}
    \hline
    \hline
 ID & $\sigma_{\Delta E(B-V)}$ & $\Delta E(B-V)_{\rm 98\% - 2\%}$ &$\sigma_{\rm (\Delta E(B-V)_{\rm B-I} - \Delta E(B-V)_{\rm U-V})}$& $R_{\rm max}$&  R$_{\rm ChM}$ \\
    &  [mag] & [mag] & [mag] & [arcmin] & [arcmin] \\
   \hline
NGC 104&    0.005 $\pm$0.001 & 0.021 $\pm$  0.001 &    0.008&   16.9& 0.0 - 16.9 \\     
NGC 288&    0.003 $\pm$0.001 & 0.010 $\pm$  0.001 &    0.003&   10.3 & 0.8 - 10.3  \\ 
NGC 1261&   0.005 $\pm$0.001 & 0.018 $\pm$  0.002 &    0.006&   5.1  & 0.0 - 5.1  \\    
NGC 1851&   0.004 $\pm$0.001 & 0.014 $\pm$  0.004 &    0.005&   8.4  & 1.2 -  8.4 \\   
NGC 1904&   0.006 $\pm$0.002 & 0.020 $\pm$  0.007 &    0.007&   6.8  & 1.5 - 6.8  \\    
NGC 2298&   0.014 $\pm$0.002 & 0.049 $\pm$  0.005 &    0.009&   4.2 & 0.9 - 4.2 \\    
NGC 2808&   0.013 $\pm$0.002 & 0.049 $\pm$  0.009 &    0.009&   6.8  & 0.0 - 6.8 \\    
NGC 3201&   0.036 $\pm$0.003 & 0.134 $\pm$  0.013 &    0.009&   7.7  & 1.2 - 7.7 \\    
NGC 4147&   0.003 $\pm$0.001 & 0.008 $\pm$  0.001 &    0.003&   3.4  & 0.0 - 3.4\\     
NGC 4372&   0.054 $\pm$0.003 & 0.187 $\pm$  0.021 &    0.016&   8.4  & $-$\\   
NGC 4590&   0.004 $\pm$0.001 & 0.013 $\pm$  0.002 &    0.004&   6.7  & 0.0 - 6.7 \\   
NGC 4833&   0.038 $\pm$0.003 & 0.138 $\pm$  0.011 &    0.015&   6.9  & 0.0 - 6.9 \\    
NGC 5024&   0.003 $\pm$0.001 & 0.013 $\pm$  0.002 &    0.005&   9.2  & $-$\\    
NGC 5053&   0.002 $\pm$0.001 & 0.008  $\pm$  0.001 &    0.001&   6.9  & 0.7 - 6.9 \\   
NGC 5272&   0.004 $\pm$0.001 & 0.015  $\pm$  0.003 &    0.008&   12.0  & 2.3 - 12.0\\    
NGC 5286&   0.023 $\pm$0.004 & 0.091  $\pm$  0.015 &    0.013&   4.2  & $-$\\    
NGC 5634&   0.006 $\pm$0.001 & 0.016  $\pm$  0.003 &    0.004&   5.0  & $-$\\
NGC 5904&   0.004 $\pm$0.001 & 0.014 $\pm$ 0.002 &    0.007&   11.9 & 2.6 - 11.9\\     
NGC 5927&   0.040 $\pm$0.004 & 0.141 $\pm$ 0.010 &    0.018&   5.1  & 1.6 - 5.1 \\  
NGC 5986&   0.020 $\pm$0.002 & 0.070 $\pm$ 0.007 &    0.017&   4.3 & $-$\\    
NGC 6101&   0.007 $\pm$0.001 & 0.026 $\pm$ 0.004 &    0.010&   7.1 & $-$\\  
NGC 6121&   0.020 $\pm$0.001 & 0.088 $\pm$ 0.008 &    0.013&   6.8  & 0.4 - 6.8\\     
NGC 6205&   0.003 $\pm$0.001 & 0.013 $\pm$ 0.001 &    0.004&   10.9 & 3.3 - 10.9\\   
NGC 6218&   0.008 $\pm$0.001 &0.029 $\pm$0.003 &   0.006&   6.7  & 1.6 - 6.7 \\    
NGC 6254&   0.018 $\pm$0.003 &0.071 $\pm$0.006 &   0.010&   10.8  &  0.0 - 10.8 \\  
NGC 6341&   0.003 $\pm$0.001 &0.011 $\pm$0.001 &   0.003&   8.4 & $-$ \\   
NGC 6366&   0.029 $\pm$0.003 &0.101 $\pm$0.007 &   0.009&   6.9  &  0.0 - 6.9 \\    
NGC 6656&   0.020 $\pm$0.003 &0.082 $\pm$0.015 &   0.019&   8.4 & $-$ \\   
NGC 6712&   0.026 $\pm$0.001 &0.079 $\pm$0.013 &   0.010&   8.5  & 0.8 - 8.5\\    
NGC 6752&   0.005 $\pm$0.001 &0.016 $\pm$0.001 &   0.003&   8.5 &  0.9 - 8.5\\    
NGC 6760&   0.050 $\pm$0.008 &0.160 $\pm$0.025 &   0.015&   3.4 &  $-$ \\  
NGC 6809&   0.006 $\pm$0.001 &0.021 $\pm$0.002 &   0.009&  10.9 & 2.1 - 10.9\\
NGC 6838&   0.019 $\pm$0.002 &0.077 $\pm$0.007 &   0.005&   3.5  &  0.0 - 3.5\\  
NGC 6934&   0.006 $\pm$0.001 &0.023 $\pm$0.004 &   0.006&   3.3  &  0.0 - 3.3\\  
NGC 6981&   0.004 $\pm$0.001 &0.016 $\pm$0.001 &   0.004&   5.3  & 0.5 - 5.3 \\
NGC 7006&   0.003 $\pm$ 0.001 &0.010 $\pm$ 0.001 &   0.002&   3.2  &  0.0 - 3.2\\    
NGC 7078&   0.009 $\pm$ 0.001 &0.032 $\pm$ 0.003 &   0.005&   8.4  & $-$ \\   
NGC 7089&   0.004 $\pm$ 0.001 &0.016 $\pm$ 0.002 &   0.003&   8.4& 1.6 - 8.4 \\    
NGC 7099&   0.003 $\pm$ 0.001 &0.010 $\pm$ 0.002 &   0.007&   8.1 & $-$\\   
NGC 7492&   0.002 $\pm$ 0.001 &0.005 $\pm$ 0.002 &   0.002&   3.4 & 0.0 - 3.4\\    
IC 4499&    0.004 $\pm$ 0.001 &0.017 $\pm$ 0.003 &   0.003&   6.6  & $-$\\   
Palomar 11&     0.018 $\pm$ 0.003 &0.054 $\pm$ 0.009 &   0.008&   2.8  & $-$\\     
Terzan 8&   0.003 $\pm$ 0.001 &0.016 $\pm$ 0.002 &   0.005&   5.1  & $-$\\
 \hline
 \hline
\end{tabular}
 \end{table*}
 %\end{landscape}
%\end{longrotatetable}

\clearpage

\begin{table}
 \caption{Fraction of 1G stars for the five GCs without previous determinations of the ChMs }
 \label{tab:tab2}
 \begin{tabular*}{\columnwidth}{@{}l@{\hspace*{50pt}}l@{\hspace*{50pt}}l@{}}
  \hline
  \hline
  ID & N$_{\rm 1G}$/N$_{\rm TOT}$& N$_{\rm stars}$\\
  \hline
  NGC\,1904 & 0.33 $\pm$ 0.03& 319\\
  NGC\,4147 & 0.36 $\pm$ 0.04&  92\\
  NGC\,6712 & 0.58 $\pm$ 0.03& 304\\
  NGC\,7006 & 0.52 $\pm$ 0.07&  55\\
  NGC\,7492 & 0.36 $\pm$ 0.04& 117\\
  \hline
  \hline
 \end{tabular*}
\end{table}

\section*{Acknowledgements}
%We thank the anonymous referee for  a constructive report that has improved the quality of the manuscript.
This work has received funding from the European Research Council (ERC) under the European Union's Horizon 2020 research innovation programme (Grant Agreement ERC-StG 2016, No 716082 'GALFOR', PI: Milone, http://progetti.dfa.unipd.it/GALFOR).
APM and ED  have been supported by MIUR under PRIN program 2017Z2HSMF (PI: Bedin)
APM and GC acknowledge support from MIUR through the FARE project R164RM93XW SEMPLICE (PI: Milone).
SJ and YWL acknowledges support from the NRF of Korea (2022R1A2C3002992, 2022R1A6A1A03053472).
%%%%%%%%%%%%%%%%%%%%%%%%%%%%%%%%%%%%%%%%%%%%%%%%%%
\section*{Data Availability}

The data underlying this article will be shared on reasonable request to the corresponding author. 
 
%The inclusion of a Data Availability Statement is a requirement for articles published in MNRAS. Data Availability Statements provide a standardised format for readers to understand the availability of data underlying the research results described in the article. The statement may refer to original data generated in the course of the study or to third-party data analysed in the article. The statement should describe and provide means of access, where possible, by linking to the data or providing the required accession numbers for the relevant databases or DOIs.



\bibliographystyle{mnras}
\bibliography{mnras_template} % if your bibtex file is called example.bib

%\begin{thebibliography}{99}
%\end{thebibliography}

%%%%%%%%%%%%%%%%%%%%%%%%%%%%%%%%%%%%%%%%%%%%%%%%%%

% Don't change these lines
\bsp	% typesetting comment
\label{lastpage}
\end{document}